\documentclass[twocolumn]{aastex62}

\usepackage{amsmath,scalerel}
\usepackage{natbib}
\bibpunct{(}{)}{;}{a}{}{,}
\usepackage{multirow}
\usepackage{graphics}
\usepackage{threeparttable}
\usepackage[varg]{txfonts}
\usepackage{xcolor}
\usepackage{bm}
\hypersetup{colorlinks,linkcolor={blue},citecolor={blue},urlcolor={blue}}

\begin{document}

\title{The Host-Galaxy Properties of Type~1 Versus Type~2 Active Galactic Nuclei}

\author{Fan Zou}
\affiliation{CAS Key Laboratory for Research in Galaxies and Cosmology, Department of Astronomy, University of Science and Technology of China, Hefei 230026, China}
\affiliation{Department of Astronomy and Astrophysics, 525 Davey Lab, The Pennsylvania State University, University Park, PA 16802, USA}
\author{Guang Yang}
\affiliation{Department of Astronomy and Astrophysics, 525 Davey Lab, The Pennsylvania State University, University Park, PA 16802, USA}
\affiliation{Institute for Gravitation and the Cosmos, 525 Davey Lab, The Pennsylvania State University, University Park, PA 16802, USA}
\author{William N. Brandt}
\affiliation{Department of Astronomy and Astrophysics, 525 Davey Lab, The Pennsylvania State University, University Park, PA 16802, USA}
\affiliation{Institute for Gravitation and the Cosmos, 525 Davey Lab, The Pennsylvania State University, University Park, PA 16802, USA}
\affiliation{Department of Physics, 104 Davey Lab, The Pennsylvania State University, University Park, PA 16802, USA}
\author{Yongquan Xue}
\affiliation{CAS Key Laboratory for Research in Galaxies and Cosmology, Department of Astronomy, University of Science and Technology of China, Hefei 230026, China}
\affiliation{School of Astronomy and Space Science, University of Science and Technology of China, Hefei 230026, China}

\email{E-mail: zf951020@mail.ustc.edu.cn, gyang206265@gmail.com}

\begin{abstract}
    The unified model of active galactic nuclei (AGNs) proposes that different AGN optical spectral types are caused by different viewing angles with respect to an obscuring ``torus''. Therefore, this model predicts that type~1 and type~2 AGNs should have similar host-galaxy properties. We investigate this prediction with 2463 \mbox{X-ray} selected AGNs in the COSMOS field. We divide our sample into type~1 and type~2 AGNs based on their spectra, morphologies, and variability. We derive their host-galaxy stellar masses ($M_\star$) through SED fitting, and find that the host $M_\star$ of type~1 AGNs tend to be slightly smaller than those of type~2 AGNs by $\Delta\overline{\mathrm{log}M_\star}\approx0.2~\mathrm{dex}$ ($\approx 4\sigma$ significance). Besides deriving star-formation rates (SFRs) from SED fitting, we also utilize far-infrared (FIR) photometry and a stacking method to obtain FIR-based SFRs. We find that the SFRs of type~1 and type~2 sources are similar once their redshifts and \mbox{X-ray} luminosities are controlled. We also investigate cosmic environment, and find that the surface number densities (sub-Mpc) and cosmic-web environments ($\approx 1\text{--}10$~Mpc) are similar for both populations. In summary, our analyses show that the host galaxies of type~1 and type~2 AGNs have similar SFR and cosmic environment in general, but the former tend to have lower $M_\star$ than the latter. The difference in $M_\star$ indicates that the AGN unification model is not strictly correct and both host galaxy and torus may contribute to the optical obscuration of AGNs.
\end{abstract}

\section{Introduction}
\label{section: intro}
Based on their optical emission lines, active galactic nuclei (AGNs) are classified as broad-line AGNs (type~1 AGNs) with at least one broad emission line ($\mathrm{FWHM}>2000~\mathrm{km}~\mathrm{s}^{-1}$) and narrow-line AGNs (type~2 AGNs) with only narrow lines (e.g., \citealt{Khachikian1974}). Conventionally, this variety of AGNs can be primarily explained by a unified scheme (e.g., \citealt{Antonucci1993Unified, Urry1995Unified, Netzer15}). The unification model argues that these two types of AGNs are intrinsically the same type of object: a supermassive black hole (SMBH) resides in the center of a galaxy; high-velocity gas in the vicinity of the black hole (the so-called broad-line region) emits broad emission lines; a thick dusty ``torus'' obscures the broad-line region for highly inclined lines of sight. For a SMBH with a mass of $10^8~M_\odot$, its typical radius is $\sim2~\mathrm{AU}$, the broad-line region is located within $\sim(6-60)\times10^{-3}~\mathrm{pc}$ of the SMBH, and the typical scale of the torus is $\sim0.5-5~\mathrm{pc}$. When viewing the system through the obscuring torus, we will observe a type~2 AGN; if the system is face-on, we can see the broad-line region directly, and the AGN will manifest as a type~1 source. The simple unified model suggests that the differences among all kinds of AGNs should only be attributed to different orientation angles relative to the line of sight and thus predicts that there should be no apparent differences between their host-galaxy properties. Some AGNs also show extreme optical emission-line variations with optical spectral types changing within months or years possibly caused by changes in the vicinity of the SMBH (e.g., \citealt{Tohline76, Penston84, Cohen86, Tran92}). If this rapid transformation in spectral types is common among AGNs, we might also expect that the host-galaxy properties would be similar for the two types of AGNs.\par
Some works argue that the unified model is not precisely correct. SMBH-galaxy co-evolution models suggest a different story from the unified model: mergers among gas-rich galaxies drive dust and gas down to the central SMBHs and fuel strong star-formation and AGN activity. During most of the rapid-accretion phase, AGN activity is obscured by large amounts of dust and gas, and thus the AGN is type~2. After the gas is consumed or swept out by stellar or AGN feedback, the AGN will gradually become unobscured (type~1). The lack of gas also leads to a decrease of star-formation activity (e.g., \citealt{Sanders1988CoEvo, Springel05CoEvo, Hopkins06CoEvo}). Overall, AGN type may be attributed to evolutionary phase. Indeed, there is increasing evidence indicating a co-evolution scheme between galaxies and SMBHs (e.g., \citealt{John13ReviwCo, Yang19}). For instance, the mass of the central SMBH is tightly correlated with bulge velocity dispersion (e.g., \citealt{Ferrarese00Msigma, Gebhardt00Msigma}) as well as bulge mass (e.g., \citealt{Magorrian1998MM, Marconi03MM}).\par
Additionally, \citet{Malkan98} found excess galactic dust in Seyfert~2 galaxies compared to Seyfert~1 galaxies by directly imaging nearby AGNs, and \citet{Maiolino95} found that the obscuration of intermediate-type Seyfert nuclei may be caused by larger-scale material instead of parsec-scale tori. Therefore, dust in host galaxies may also be responsible for the obscuration of AGNs (e.g., \citealt{Matt00, Netzer15}), and thus tori might not be necessary to provide obscuration for all type~2 AGNs. This scenario predicts that the host galaxies of type~2 AGNs will likely be more massive than those of type~1 AGNs, since dust is generally more abundant in more massive galaxies (e.g., \citealt{Katherine17}).\par
Given the different physical scenarios between the unified model and other models, at least two critical questions can be raised: whether the obscuring material is in the form of a parsec-scale torus rather than galaxy-scale gas, as assumed in the unified model; and whether type~1 and type~2 sources are same objects intrinsically.\par
Observationally, there is hardly a consensus concerning the difference between the host-galaxy properties of different AGN types in both local-universe studies and distant-universe studies. In the nearby universe, \citet{Trump13} found that type~1 AGN hosts are less likely to reside in the red sequence and have a younger stellar population, while \citet{Kauffmann03} argued that there is not any difference in stellar contents if AGN luminosities are sufficiently high. Conclusions about cosmic environments in the local universe are also controversial. On a large scale ($\gtrsim1~\mathrm{Mpc}$), \citet{Powell18} reported that nearby obscured AGNs reside in more massive halos than unobscured AGNs, but \citet{Jiang16} found that their halo masses are similar. Besides, on a small scale (sub-Mpc), obscured sources usually reside in a denser environment and have more neighbor galaxies (e.g., \citealt{Jiang16, Powell18}). Type~1 and type~2 AGN host galaxies might have different morphologies \citep{Maiolino1997}. Their neighbor galaxies might also have different properties such as color, AGN activity, and morphology \citep{Villarroel14}.\par
In the distant universe, the conclusions are also controversial. \citet{Merloni14} found that galaxies hosting either obscured or unobscured AGNs have the same mean SFRs at high redshift, but some works argued that type~2 hosts have higher SFRs than type~1 hosts either in the distant universe (e.g., \citealt{Bornancini18}) or the local universe (e.g., \citealt{Villarroel17}). In the case of $M_\star$ at high redshift, some works showed that there is no significant difference in $M_\star$ between obscured and unobscured AGN host galaxies (e.g., \citealt{Merloni14, Bornancini18}), but some found a strong correlation between the \mbox{X-ray} obscuration and $M_\star$ (e.g., \citealt{Lanzuisi17}). As for the environment, \citet{DiPompeo14} found that obscured AGNs display a higher clustering amplitude, and thus reside in higher mass dark-matter halos, but \citet{Bornancini18} found that the difference in the environment only exists on a small scale ($\lesssim0.1~\mathrm{Mpc}$).\par
In this work, we investigate whether different types of distant AGNs have different host $M_\star$, SFRs, and environments, including both local (on sub-Mpc scales) and global (on $1-10$~Mpc scales) environments. Our work is based on the COSMOS-Legacy Survey \citep{Civano16COSMOS}, which allows detailed analyses and significant improvements on this topic. Also, COSMOS is covered by an intensive investment of spectroscopic and multi-wavelength observations (e.g., \citealt{Lilly+09, COSMOS15}), allowing accurate estimations of the host $M_\star$ and SFRs. Thanks to the latest environment catalog in the COSMOS field \citep{Yang18Env}, our work is the first one to compare the surface number density and cosmic-web environments of type~1 and type~2 AGNs at high redshift.\par
This paper is organized as follows. In Section~\ref{section: data}, our data and samples are described. In Section~\ref{section: parameter}, we derive necessary physical parameters including AGN type, \mbox{X-ray} luminosity, stellar mass, SFR, and environmental parameters. In Section~\ref{section: result}, we present our results. Finally, we summarize and discuss our results in Section~\ref{section: summary}. Throughout this paper, we adopt a flat $\Lambda\mathrm{CDM}$ cosmology with $H_0=70~\mathrm{km}~\mathrm{s}^{-1}~\mathrm{Mpc}^{-1}$, $\Omega_\Lambda=0.73$ and $\Omega_M=0.27$. Errors are given at a $1\sigma$ (68\%) confidence level. Units of $L_\mathrm{X}$, $M_\star$, and SFR are ``$\mathrm{erg}~\mathrm{s}^{-1}$'', solar mass ``$M_\odot$'' and ``$M_\odot~\mathrm{yr}^{-1}$'', respectively. Following standard practice, we adopt $2\sigma$ (\emph{p}-value = 0.05) as the threshold for a ``statistically significant'' difference in host-galaxy properties, i.e., we consider two quantities as consistent if the significance of the difference is smaller than $2\sigma$. When multiple trials are being used in a hypothesis test concurrently but only one of them exceeds $2\sigma$, we apply the Bonferroni correction \citep{Bonferroni36} to adjust the required significance level corresponding to \emph{p}-value = 0.05.

\section{Data and Sample Selection}
\label{section: data}
\subsection{Sample Selection}
\label{section: sample selection}
Our sample is drawn from \mbox{X-ray} detected sources in the COSMOS-Legacy Survey \citep{Civano16COSMOS}. This survey reaches a flux limit three times deeper than the \emph{XMM}-COSMOS Survey \citep{Cappelluti09, Brusa10}, and it also covers a larger area ($\sim2~\mathrm{deg}^2$) than the C-COSMOS Survey \citep{Elvis09, Civano12}. Therefore, the COSMOS-Legacy Survey allows us to obtain a larger and more complete AGN sample compared to previous works (e.g., \citealt{Merloni14}). The COSMOS-Legacy Survey samples most of cosmic accretion power (e.g., \citealt{Aird15, Yang18Env}). In this sense, the sources studied in this work are typical AGNs in the distant universe. The survey reaches $\approx10$ times below the knee luminosity of the \mbox{X-ray} luminosity function at $z\approx1.5-3$, the cosmic epoch when mergers are most important to galaxy evolution (e.g., \citealt{Conselice14}). In this epoch, the aforementioned merger-driven coevolution scenario might be crucial in shaping SMBH/galaxy growth. Therefore, the COSMOS-Legacy Survey is ideal to probe the possible differences between type~1 and type~2 sources.\par
In addition, dilution by host-galaxy starlight is low in the \mbox{X-ray} band, thus allowing us to construct pure AGN samples even down to modest luminosities (e.g., \citealt{Brandt15Review, XueReview} and references therein). The \mbox{X-ray} catalog with optical and infrared (IR) identifications in COSMOS is available in \citet{Marchesi16COSMOS}. The high angular resolution of \emph{Chandra} ($\sim0.5''$) in the COSMOS-Legacy Survey enables reliable cross-matching between the \mbox{X-ray} catalog and IR-to-UV catalogs \citep{Marchesi16COSMOS}.\par
We select 3744 sources that are not stars with non-zero redshift from \citet{Marchesi16COSMOS}. Then we remove sources outside of the UltraVISTA area or masked in optical broad-bands \citep{Capak07COSMOS} to ensure accurate SED fitting results, following previous works (e.g., \citealt{Yang18Env, Yang18_2, Kashino18}). The UltraVISTA area has deep near-infrared (NIR) imaging data that are essential in estimating photometric redshifts and $M_\star$, and the masked regions do not have accurate photometry because they are located beside bright sources or have bad pixels. Indeed, we find that the photometric-redshift quality\footnote{As measured by $\sigma_\mathrm{nmad}=1.48\times\mathrm{median}\{|z_\mathrm{p}-z_\mathrm{s}|/(1+z_\mathrm{s})\}$, where $z_\mathrm{p}$ and $z_\mathrm{s}$ are photometric and spectroscopic redshifts, respectively.} in the masked regions is three times worse than that in the unmasked region, indicating that removing the sources is necessary. We have also tested including the masked sources with spectroscopic redshifts when analyzing FIR-based SFRs, and find that the results in Section \ref{section: SFR} do not change qualitatively, indicating that removing these sources does not cause significant biases. There are 2463 sources left in total after removing the masked sources. We adopt the redshifts in \citet{Marchesi16COSMOS}, and 1434 sources have reliable spectroscopic-redshift measurements. We derive physical parameters for our sources below and release them in a source catalog. A part of our source catalog is displayed in Table~\ref{CatTable}, and the full version is available in the online supplementary materials.

\begin{table*}[hpbt]
\caption{Source catalog.}
\label{CatTable}
\resizebox{\hsize}{!}{
\centering
\begin{threeparttable}
\begin{tabular}{ccccccccccccc}
\hline
\hline
RA & Dec & Redshift & z\_type & Class & $\mathrm{log}L_\mathrm{X}$ & $\mathrm{log}M_\star$ & $\mathrm{log}M_{\star,~\mathrm{BT1}}$ & $\mathrm{logSFR}_\mathrm{SED}$ & $\mathrm{logSFR}_\mathrm{FIR}$ & $\mathrm{log}(1+\delta)$ & Web\\
(degree) & (degree) & & & & ($\mathrm{erg}~\mathrm{s}^{-1}$) & ($M_\odot$) & ($M_\odot$) & ($M_\odot~\mathrm{yr}^{-1}$) & ($M_\odot~\mathrm{yr}^{-1}$) & &\\
(1) & (2) & (3) & (4) & (5) & (6) & (7) & (8) & (9) & (10) & (11) & (12)\\
\hline
149.4116608 & 2.4103652 & 1.063 & spec & ST2 & 43.80 & 9.04 & $-99.0$ & 1.04 & $-99.0$ & $-0.192$ & 1\\
149.4166403 & 2.6936599 & 0.859 & phot & QT2 & 43.53 & 10.65 & $-99.0$ & 1.17 & $-99.0$ & $-0.081$ & 1\\
149.4179818 & 2.2657195 & 2.710 & phot & QT2 & 44.57 & 10.10 & $-99.0$ & 2.17 & $-99.0$ & $-99.0$ & $-99$\\
149.4199481 & 2.0355366 & 1.480 & spec & ST1 & 44.72 & 10.60 & 10.20 & 2.27 & 2.06 & 0.302 & 1\\
149.4219104 & 2.2080634 & 2.930 & phot & QT2 & 44.48 & 10.76 & $-99.0$ & 1.06 & $-99.0$ & $-0.073$ & 2\\
149.4231131 & 2.4447908 & 1.741 & phot & QT1 & $44.71$ & 11.08 & 11.21 & 1.26 & $-99.0$ & $-99.0$ & $-99$\\
149.4244121 & 1.7302479 & 1.535 & phot & QT2 & 43.53 & 11.51 & $-99.0$ & 1.50 & $-99.0$ & $-0.008$ & 1\\
149.4250437 & 2.7581464 & 2.750 & phot & QT1 & 44.85 & 10.38 & 10.41 & 3.40 & $-99.0$ & $-99.0$ & $-99$\\
149.4297092 & 2.3892849 & 2.133 & spec & ST2 & 44.57 & 10.90 & $-99.0$ & 1.84 & 2.47 & 0.084 & 1\\
149.4327390 & 2.3775369 & 1.852 & spec & ST2 & 44.13 & 11.27 & $-99.0$ & 1.96 & 2.18 & 0.026 & 1\\
\hline
\hline
\end{tabular}
\begin{tablenotes}
\small
\item
\emph{Notes.} The full version of this catalog is available in the online supplementary materials. The table is sorted in ascending order of RA. (1) and (2) J2000 coordinates. (3) Redshift. (4) Redshift type. ``spec'' and ``phot'' indicate that the redshifts are spectroscopic and photometric redshifts, respectively. (5) AGN classification. There are four kinds of classes: spectroscopic type~1 (ST1), spectroscopic type~2 (ST2), \emph{QSOV} type~1 (QT1), and \emph{QSOV} type~2 (QT2). See Section~\ref{section: classification} for more details of the classifications. (6) \mbox{X-ray} luminosity (rest-frame $2-10~\mathrm{keV}$ luminosity; Section~\ref{section: LX}). (7) and (8) $M_\star$. The subscript ``BT1'' of the eighth column indicates use of the type~1 template in \citet{Buat15} (Section~\ref{section: Mstar}); (9) SED-based SFR. For the ST2 and QT2 samples, $\mathrm{log}M_{\star,~\mathrm{BT1}}$ are set to $-99.0$. (10) FIR-based SFR (Section~\ref{section: SFR}). (11) Dimensionless overdensity parameter (Section~\ref{section: DefEnv}). (12) Cosmic-web environment (0: cluster; 1: filament; 2: field; Section~\ref{section: DefEnv}). Unavailable values are set to $-99$.
\end{tablenotes}
\end{threeparttable}
}
\end{table*}

\subsection{NUV to NIR Photometry}
\label{section: COSMOS2015}
Photometric data from the near-ultraviolet (NUV) to NIR are from the COSMOS2015 catalog \citep{COSMOS15}, the latest catalog containing multi-wavelength photometry in COSMOS. To ensure that fluxes are consistent over the full wavelength range, we correct the fluxes using the following formulae:
\begin{align}
f_{i, j}^{\mathrm{TOT},0}&=f_{i, j}^\mathrm{APER3}10^{-0.4o_i}, \mathrm{and}\\
f_{i, j}^\mathrm{TOT}&=f_{i, j}^{\mathrm{TOT},0}10^{0.4(EBV_iF_j+s_j)},
\end{align}
where $f_{i, j}^\mathrm{APER3}$ is the $3''$ diameter aperture flux; $f_{i, j}^{\mathrm{TOT},0}$ is the uncorrected total flux of the source; $f_{i, j}^\mathrm{TOT}$ is the corrected total flux and is adopted when performing SED fitting; $i$ is the object identifier, and $j$ is the filter identifier; $o_i$ is a single offset allowing for the conversion from aperture magnitude to total magnitude; $EBV_iF_j$ represents the foreground Galactic extinction, where $EBV_i$ is the reddenning value of the object and $F_j$ is the extinction factor; $s_j$ is a systematic offset. All these correction factors are available in the COSMOS2015 catalog. Formulae to correct flux errors are similar, except that the errors should be multiplied by another factor representing the effect of correlated noise, which is also available in the COSMOS2015 catalog.

\subsection{FIR Photometry}
\label{section: FIR}
FIR photometry is vital to constrain SFR, especially for type~1 host galaxies with low SFRs because optical-UV emission from the AGN component may be mistakenly attributed to starlight without FIR photometry (Section~\ref{section: sedfitting}), and thus even upper limits on FIR fluxes are worth utilizing. $100~\mu\mathrm{m}$ and $160~\mu\mathrm{m}$ fluxes are from the $24~\mu\mathrm{m}$ prior catalog in the public data release of the \emph{Herschel}-Photodetector Array Camera and Spectrometer (PACS) Evolution Probe (PEP; \citealt{PEPLutz}). $250~\mu\mathrm{m}$, $350~\mu\mathrm{m}$, and $500~\mu\mathrm{m}$ fluxes, which were observed using the \emph{Herschel}-Spectral and Photometric Imaging Receiver (SPIRE), are from the XID+ catalog of the \emph{Herschel} Multi-Tiered Extragalactic Survey (HerMES; \citealt{Oliver12, HerMES}). The XID+ catalog also uses $24~\mu\mathrm{m}$ detected sources as a prior list for extracting fluxes. We adopt $2''$ as our matching radius and use $24~\mu\mathrm{m}$ positions in these two catalogs when matching with other catalogs because $24~\mu\mathrm{m}$ positions are more accurate than positions at longer wavelengths.\par
For the XID+ catalog, we take the maximum of (84th--50th percentile) and (50th--16th percentile) of marginalized flux probability distributions as estimated errors $S_\mathrm{err}$ assuming Gaussian uncertainties for each band. We remove sources with signal-to-noise ratio $S/S_\mathrm{err}<3$. Fluxes of all the selected sources are above 5 mJy at $250~\mu\mathrm{m}$, 5 mJy at $350~\mu\mathrm{m}$, and 7 mJy at $500~\mu\mathrm{m}$, and thus the Gaussian approximation to uncertainties is valid according to the documentation for the COSMOS-XID+ catalog of HerMES.\par
For sources undetected at $100~\mu\mathrm{m}$ or $160~\mu\mathrm{m}$, we use a method similar to that in \citet{Stanley15Upplim} to estimate their flux upper limits in each band. We perform aperture photometry at about 300 positions around a non-detection point in residual maps, taking the 68.4th percentile of the distribution of the measured flux densities as the $1\sigma$ upper limit on the non-detection and multiply the value by 3 to get the $3\sigma$ flux upper limit. We do not derive flux upper limits at $250~\mu\mathrm{m}$, $350~\mu\mathrm{m}$, and $500~\mu\mathrm{m}$ because there are not public residual maps for those bands and the SPIRE maps are confusion-limited, making estimating background fluxes only using the simple aperture photometry method infeasible.\par

\section{Derivation of Physical Parameters}
\label{section: parameter}
\subsection{Classification of Type~1 and Type~2 Sources}
\label{section: classification}
\citet{Marchesi16COSMOS} compiled and fitted the available optical/NIR spectra for the X-ray AGNs in the COSMOS-Legacy Survey. They classified a source as type 1 if it had at least one broad emission line with $\mathrm{FWHM}>2000~\mathrm{km~s}^{-1}$; otherwise, they classified it as type 2. We adopt their classifications\footnote{Since we do not have access to most of the AGN spectra, we cannot fit the emission lines and do the classifications by ourselves.} and obtain 1162 sources in total, including 326 spectroscopic type~1 sources and 836 spectroscopic type~2 sources. This case is labeled as ``Case~1''.\par
Although spectroscopic classifications are reliable, they inevitably suffer from incompleteness issues, as 1301 sources among our X-ray AGNs do not have spectroscopic classifications. To address such problems, we classify AGNs without spectroscopic classifications in \citet{Marchesi16COSMOS} based on their host-galaxy morphologies and optical variability. For point-like sources, the host galaxies do not contribute much to the observed emission, and thus the nuclei are unlikely obscured; for highly variable sources, their emissions are also dominated by AGNs because host galaxies generally do not have detectable variability \citep{Salvato09}. These morphology and variability classification schemes are widely adopted in the literature (e.g., \citealt{Merloni14, Salvato11}), and we also adopt these classifications for sources without spectra. We will also test the reliability of the classifications in this section.\par
We use the morphology parameter in \citet{Leauthaud07} to identify point sources. They selected point sources in deep COSMOS \emph{Hubble Space Telescope} (\emph{HST})/Advanced Camera for Surveys (ACS) F814W images \citep{Koekemoer07}. The optical variability measurements are from \citet{Salvato11}, who measured the variability in the C-COSMOS field and the \emph{XMM}-COSMOS field. Following \citet{Salvato11}, we define sources with $\mathrm{VAR}>0.25~\mathrm{mag}$ as varying sources, where VAR is defined in \citet{Salvato09}. We will call the method that selects type~1 AGNs with point-like host morphology or $\mathrm{VAR}>0.25~\mathrm{mag}$ as the \emph{QSOV} (short for point-like or varying) method hereafter.
Based on the \emph{QSOV} method, an additional \emph{QSOV} type~1 sample of 355 sources is selected among 1301 AGNs that do not have information about spectral type, with a fraction of 27.3\%. Consequently, there are 946 sources left, and we will call this sample the \emph{QSOV} type~2 sample.\par

\begin{figure}[htb!]
\resizebox{\hsize}{!}{
\includegraphics{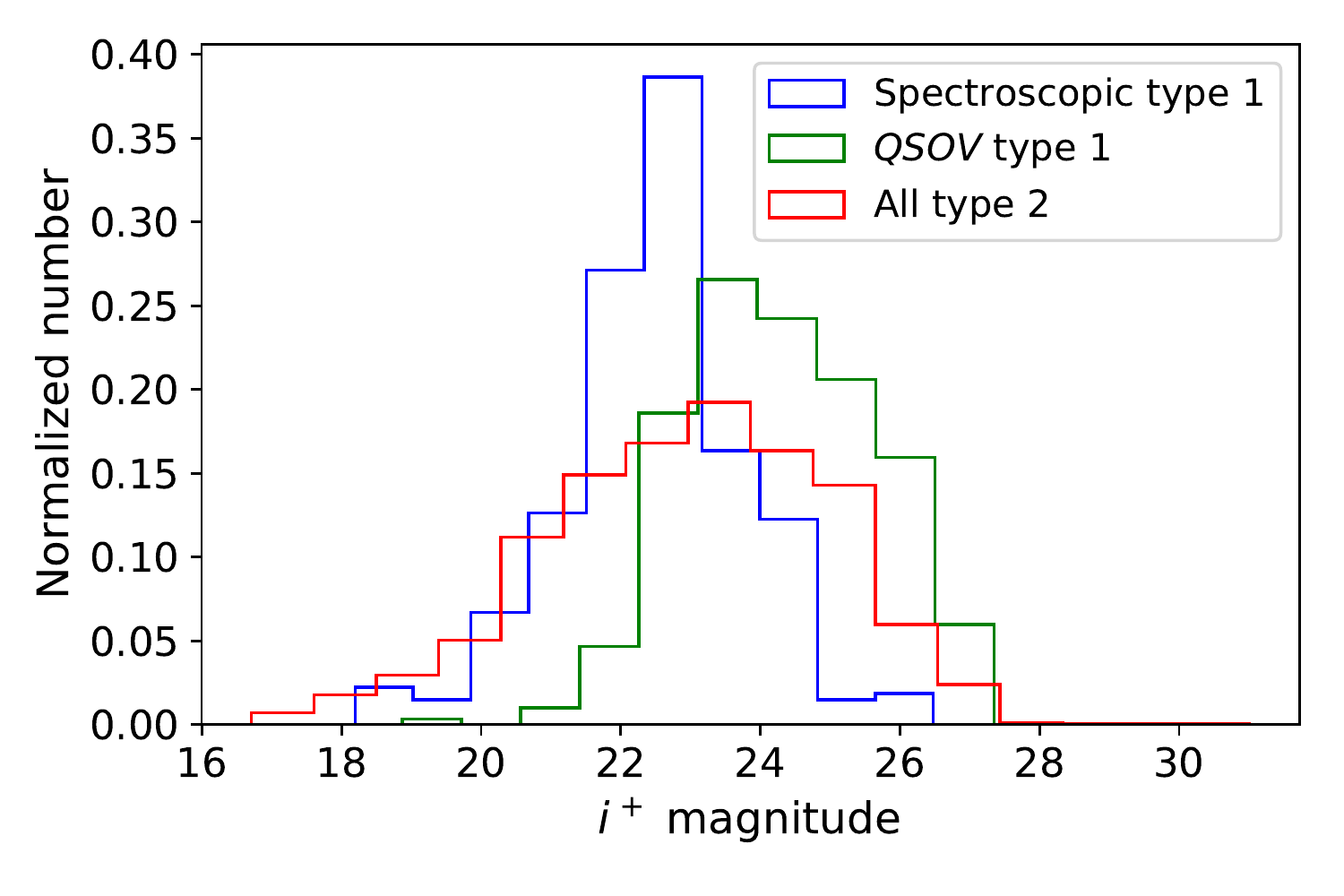}
}
\caption{Distributions of $i^+$ magnitudes of the spectroscopic type~1 sample (blue), the \emph{QSOV} type~1 sample (green), and all the type~2 sources (red). The histograms are normalized so that the sums of the histograms are 1. The \emph{QSOV} type~1 sample is fainter than other samples.\label{mag_N}}
\end{figure}

As a reliability check of the \emph{QSOV} method, we apply it to the spectroscopic type~1 sample containing 326 sources, and find that there 235 sources satisfying the criterion, with a fraction of 72.5\%. The high fraction indicates that the method is reliable for selecting type~1 AGNs. Indeed, we find that the \emph{QSOV} type~1 sample tends to have bluer rest-frame $U-V$ colors than the type~2 samples, indicating the presence of type~1 AGNs. We also plot the distributions of the magnitudes in the $i^+$ band for the spectroscopic type~1 sample, the \emph{QSOV} type~1 sample, and all the type~2 samples including the spectroscopic type~2 sample and the \emph{QSOV} type~2 sample in Fig.~\ref{mag_N}. The figure shows that the \emph{QSOV} type~1 sources are generally fainter than other sources. Therefore, it is possible that their optical faintness prevents robust spectroscopic identification of their type 1 nature.\par

We add the \emph{QSOV} type~1 and type~2 sources into the type~1 and type~2 samples in a new case, which is labeled as ``Case~2'' hereafter. Case~2 is the same as the classification scheme of \citet{Merloni14}. Compared to Case~1, the samples in Case~2 are more complete. We summarize the properties of each case in Table~\ref{CaseTable}, and both cases can be used together to assess any sensitivity of our conclusions to AGN type classification issues.

\begin{table}[hpbt]
\centering
\caption{
Compositions of type~1 and type~2 samples in each Case.
}
\label{CaseTable}
\begin{threeparttable}
\begin{tabular}{c|cc}
\hline
\hline
& Type~1 & Type~2\\
\hline
Case~1 (1162) & ST1 (326) & ST2 (836)\\
Case~2 (2463) & ST1+QT1 (681) & ST2+QT2 (1782)\\
\hline
\hline
\end{tabular}
\begin{tablenotes}
\small
\item
\emph{Notes.} The table shows the compositions of the samples in each Case. ``ST1'', ``ST2'', ``QT1'', and ``QT2'' are short for the ``spectroscopic type~1'', ``spectroscopic type~2'', ``\emph{QSOV} type~1'', and ``\emph{QSOV} type~2'' samples, respectively. The sample sizes are shown in the parentheses.
\end{tablenotes}
\end{threeparttable}
\end{table}

Below we will analyze all the parameters in both cases independently. We find that the results are similar, and thus our results should be robust. Besides, we also try applying a magnitude cut (F814W magnitude $<24$) and find that our conclusions are not affected when the magnitude cut is applied, which further improves the reliability. However, since the magnitude cut reduces our sample size significantly and may cause bias toward higher $M_\star$ and SFR, we do not show the results with the magnitude cut below.

\subsection{\mbox{X-ray} Luminosity}
\label{section: LX}
We calculate $2-10~\mathrm{keV}$ luminosity corrected for absorption, $L_\mathrm{X}$, using the following formula assuming a power law spectrum with photon index $\Gamma=1.8$:
\begin{align}
L_\mathrm{X}=4\pi D_L^2(1+z)^{\Gamma-2}\frac{10^{2-\Gamma}-2^{2-\Gamma}}{E_2^{2-\Gamma}-E_1^{2-\Gamma}}f_{E_1\sim E_2}\eta^{-1}(i,E_1,E_2),
\end{align}
where $f_{E_1\sim E_2}$ is the flux between $E_1$ and $E_2$ in the observed-frame, $E_1$ and $E_2$ are in keV, $D_L$ is the luminosity distance at redshift $z$, and $\eta(i,E_1,E_2)$ is a factor to correct absorption between $E_1$ and $E_2$ for source $i$. $\eta(i,E_1,E_2)$ is available in \citet{Marchesi16COSMOS}. We use hard-band ($E_1=2~\mathrm{keV}$, $E_2=7~\mathrm{keV}$) fluxes for sources detected in the hard band (74.3\%), full-band ($E_1=0.5~\mathrm{keV}$, $E_2=7~\mathrm{keV}$) fluxes for sources detected in the full band and undetected in the hard band (25.0\%), and soft-band ($E_1=0.5~\mathrm{keV}$, $E_2=2~\mathrm{keV}$) fluxes for sources only detected in the soft band (0.7\%) as $f_{E_1\sim E_2}$ to calculate $L_\mathrm{X}$. Such a prioritization is chosen to minimize the effect of absorption which is less significant for harder \mbox{X-ray} photons.\par
\begin{figure}[htb!]
\resizebox{\hsize}{!}{
\includegraphics{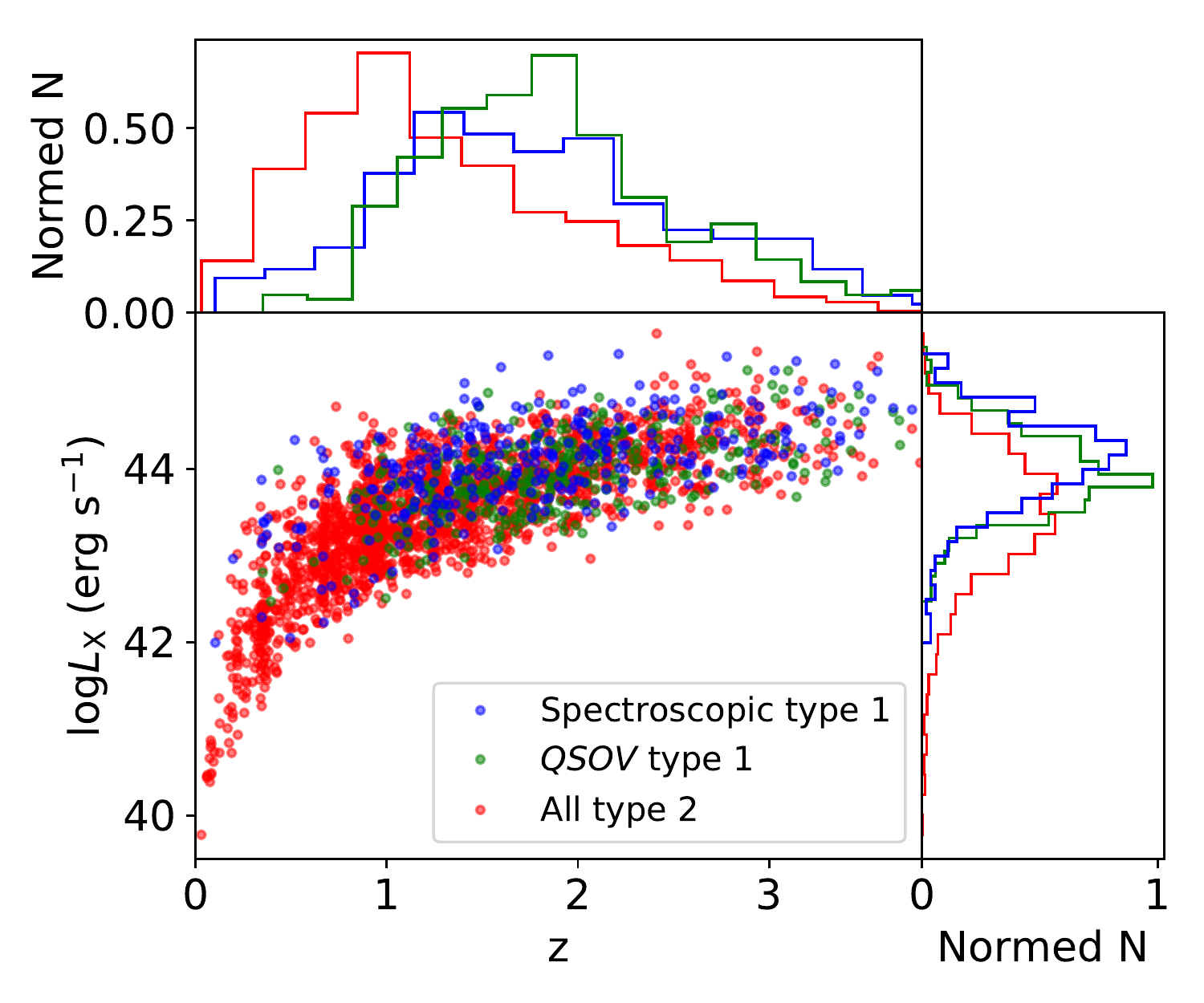}
}
\caption{Spectroscopic type~1 (blue), \emph{QSOV} type~1 (green), and type~2 (red) AGNs in our sample in the $L_\mathrm{X}-z$ plane. Distributions of $z$ and $L_\mathrm{X}$ are also shown in the \emph{top} panel and the \emph{right} panel, respectively.\label{zLxFig}}
\end{figure}

Fig.~\ref{zLxFig} displays our samples in the $L_\mathrm{X}-z$ plane. Spectroscopic type~1 AGNs and \emph{QSOV} type~1 AGNs tend to have higher $L_\mathrm{X}$ compared to type~2 AGNs because the fraction of unobscured AGN increases with $L_\mathrm{X}$ (e.g., \citealt{Merloni14}). Besides, \emph{QSOV} type~1 AGNs have lower $L_\mathrm{X}$ than spectroscopic type~1 AGNs because the \emph{QSOV} type~1 sample is generally fainter.

\subsection{SED fitting}
\label{section: sedfitting}
We use the Bayesian-based code \texttt{CIGALE}\footnote{https://cigale.lam.fr} \citep{cigaleNoll, cigaleSerra, Boquien18} to perform SED fitting. The median number of photometric bands in our SEDs is 32, and 84\% of our sources have over 30 photometric bands available. The relatively large number of bands increases our fitting reliability. We adopt an exponentially decreasing SFR model for the star-formation history (SFH). Stellar templates are from models in \citet{bc03} with metallicity of 0.0001, 0.0004, 0.004, 0.008, 0.02, or 0.05, and a Chabrier initial mass function \citep{ChabrierIMF} is assumed when measuring $M_\star$. Dust attenuation is assumed to follow a Calzetti extinction law \citep{dustatt_calzleit} allowing $E(B-V)$ of the young stellar population to vary between 0 and 1 with a step of 0.1, and $E(B-V)$ of the old stellar population is assumed to be scaled down by 0.44 compared to that of the young one. Nebular and dust emission are also implemented in \texttt{CIGALE} \citep{Draine07, cigaleNoll}. As for the AGN component, we use AGN templates in \citet{fritz2006} allowing $f_\mathrm{AGN}$ to vary between 0 and 1 with a step of 0.05, where $f_\mathrm{AGN}$ is the fractional contribution of the AGN to the total IR luminosity. The angle between the line of sight and accretion disk, $\psi$, is set to be $0^\circ$ for type~1 sources and $90^\circ$ for type~2 sources. The optical depth $\tau$ at $9.7~\mu\mathrm{m}$ is fixed to 6.0. Our settings are similar to those of many other works (e.g., \citealt{COSMOS15, Yang18_2}).\par
\begin{figure*}[htb!]
\resizebox{\hsize}{!}{
\includegraphics{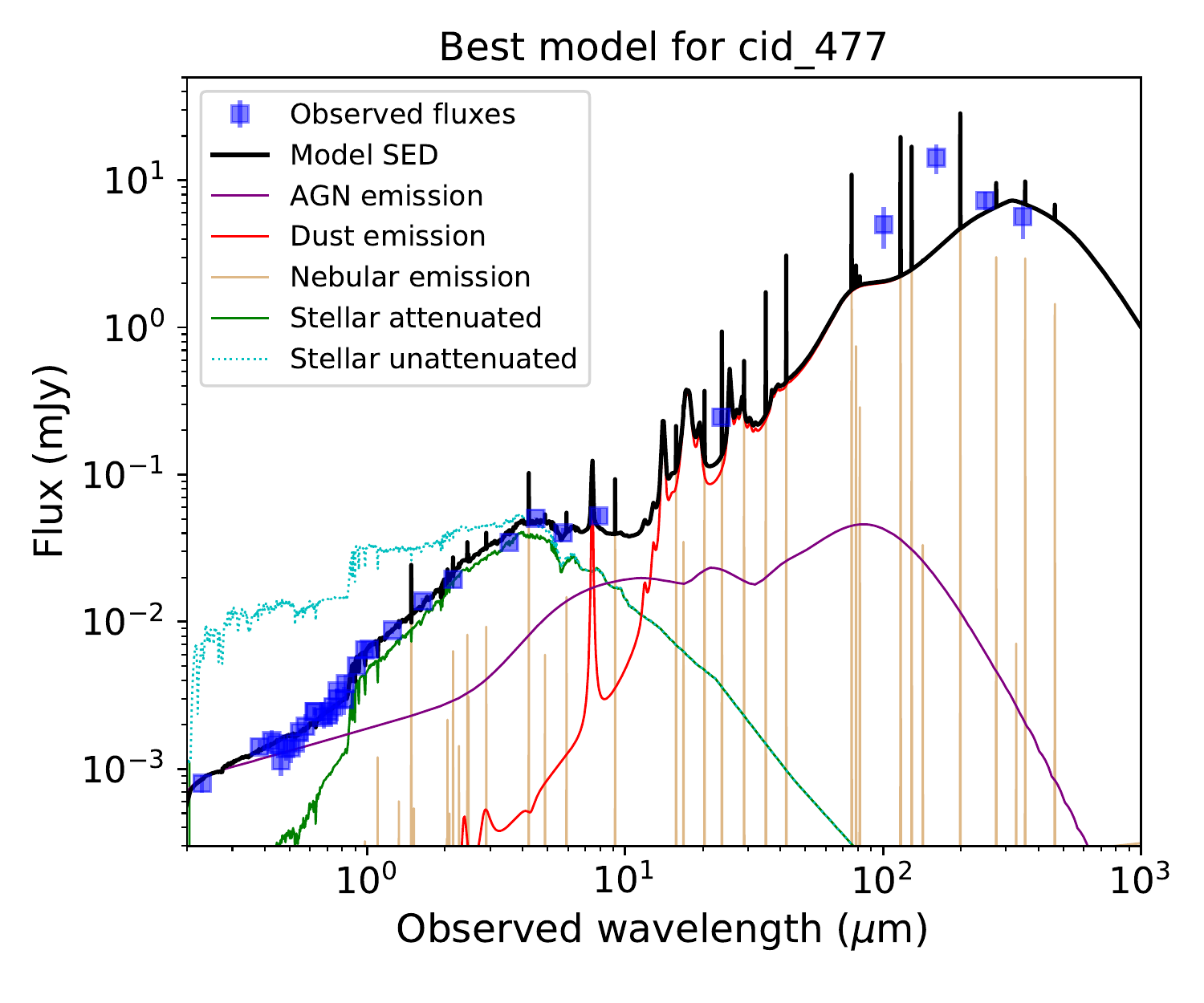}
\includegraphics{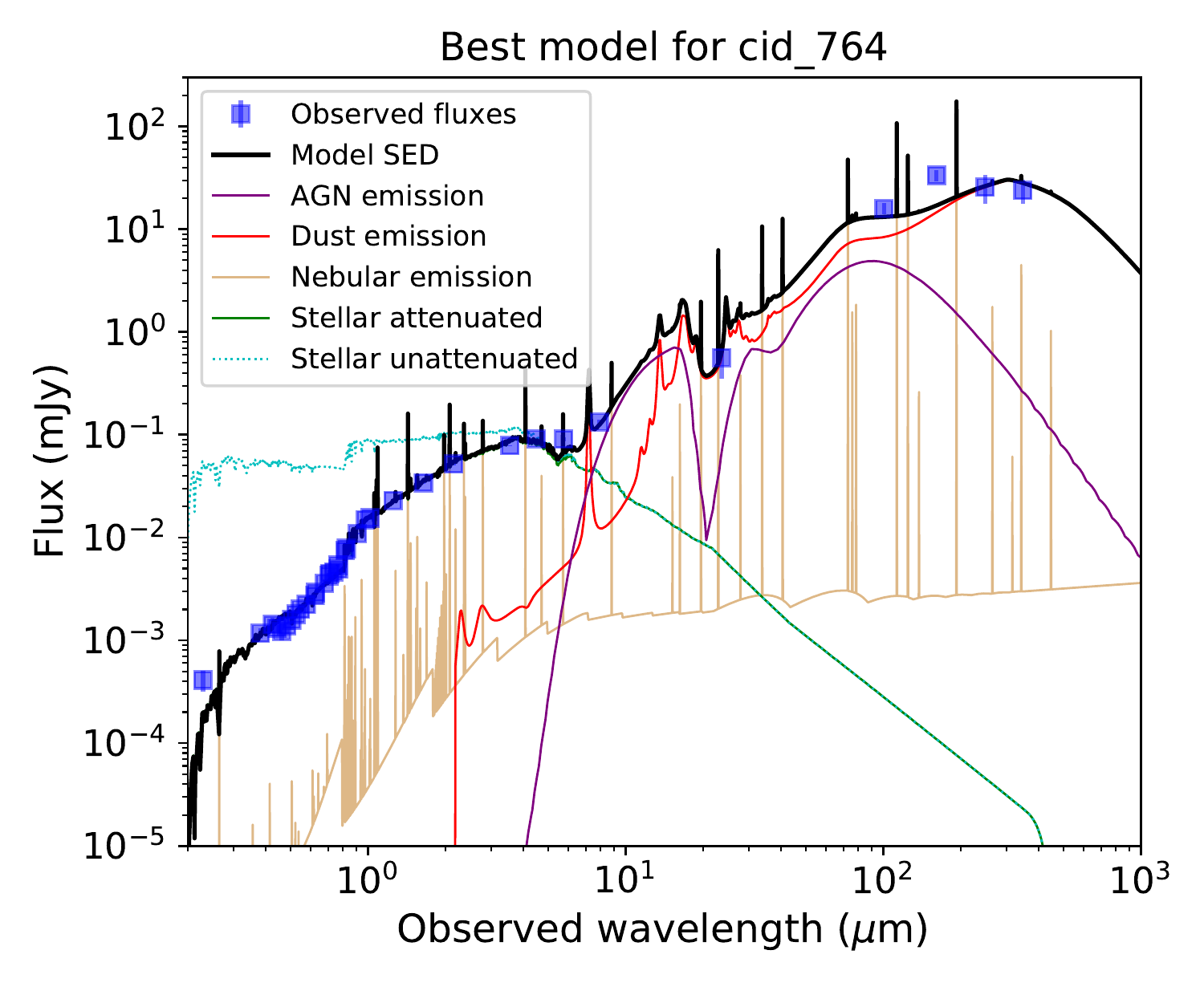}
}
\caption{Example SEDs of type~1 (\emph{left}, with 35 observed fluxes) and type~2 (\emph{right}, with 35 observed fluxes) AGNs. Blue square points are observed fluxes, and black solid curves are the best-fit model SEDs for the data. The SEDs are decomposed into AGN components (purple) and galaxy components, and the galaxy components are further decomposed into the stellar (green), nebular (brown), and dust (red) components. Dust emission dominates in the FIR band.\label{typicalSEDfig}
}
\end{figure*}

We obtain SFR and $M_\star$ from the output results. Fig.~\ref{typicalSEDfig} displays example SEDs for type~1 and type~2 AGNs. The total SEDs are decomposed into the AGN components and galaxy components, with the latter being further decomposed into the stellar, nebular, and dust components.\par
\texttt{CIGALE} adopts energy conservation for galaxies, i.e., all the energy absorbed in the UV-to-optical band is assumed to be reemitted by dust. The figure shows that dust emission dominates in FIR band. Therefore, we can have a tight constraint on the amount of dust as well as its attenuation based on the FIR data. Consequently, a better constraint on unattenuated stellar emission in the UV-to-optical band is obtained, from which \texttt{CIGALE} can derive SFR. Thus FIR photometry is very important for SFR calculation, especially for type~1 hosts because AGNs significantly contribute to their UV-to-optical emission and the decomposition of the AGN component and galaxy component is difficult without FIR photometry. Indeed, SFR can be roughly estimated from single-band FIR photometry assuming typical SEDs (e.g., \citealt{calSFR, GuangSFR}).\par
In the rest-frame NIR band, the galaxy component often contributes more than the AGN component because the emission from the AGN is weak while stellar emission peaks in that band. Such contrast improves the reliability of measured $M_\star$ since NIR flux is critical in deriving $M_\star$ (e.g., \citealt{Ciesla15, Yang18_2}).

\subsection{Cosmic Environment}
\label{section: DefEnv}
We also investigate possible differences between the cosmic environments of type~1 hosts and type~2 hosts. We probe both host-galaxy local (sub-Mpc) and global ($\sim1-10~\mathrm{Mpc}$) environment based on the latest work on the cosmic environment in COSMOS \citep{Yang18Env}, who utilized a new technique to construct a measurement of the density field and the cosmic web up to $z=3$ for COSMOS. To assess the local environment, they define a dimensionless over-density parameter for each source:
\begin{align}
1+\delta=\frac{\Sigma}{\Sigma_\mathrm{median}},
\end{align}
where $\Sigma$ is surface number density, and $\Sigma_\mathrm{median}$ is the median value of $\Sigma$ within $z\pm0.2$ at redshift $z$. They also mapped the sources to the field, filaments, or clusters of the cosmic web to assess the global environments. However, cluster signals are often dominated by noise at high redshift ($z\gtrsim1.2$) when clusters are still forming and are usually in the form of proto-clusters, and hence they only assign the global environment of sources into field and filament categories above $z=1.2$. Generally, the over-density ($1+\delta$) rises from the field to cluster environments, but there are substantial overlapping ranges of the over-density for sources in different cosmic web categories.

\section{Results}
\label{section: result}
\subsection{Stellar Mass}
\label{section: Mstar}
Since the $z$ and $L_\mathrm{X}$ distributions are different for different types of AGNs (Fig.~\ref{zLxFig}), we need to control for them to avoid a possible difference in $M_\star$ caused by different $z$ or $L_\mathrm{X}$. For example, AGNs tend to be found in massive galaxies with $\mathrm{log}M_\star\gtrsim10.5$ at $z\gtrsim1$ (e.g., \citealt{Yang18_2}), while AGN host galaxies are less massive at $z\sim0$ (e.g., \citealt{Heckman14}). Therefore, this redshift-related bias may affect our results if redshifts are not not carefully controlled in our analysis. We divide the $\mathrm{log}L_\mathrm{X}-z$ plane into a grid with $\Delta z=0.2$ and $\Delta\mathrm{log}L_\mathrm{X}=0.3~\mathrm{dex}$. Denoting the numbers of type~1 hosts and type~2 hosts in a grid element G as $N_\mathrm{G,1}$ and $N_\mathrm{G,2}$, respectively, we randomly select $\mathrm{min}\{N_\mathrm{B,1},N_\mathrm{B,2}\}$ type~1 sources as well as the same number of type~2 sources in the considered grid element. After repeating the procedure in each bin, we can construct new type~1 and type~2 samples with similar distributions of $z$ and $L_\mathrm{X}$.\par
\begin{figure*}[htb!]
\resizebox{\hsize}{!}{
\includegraphics{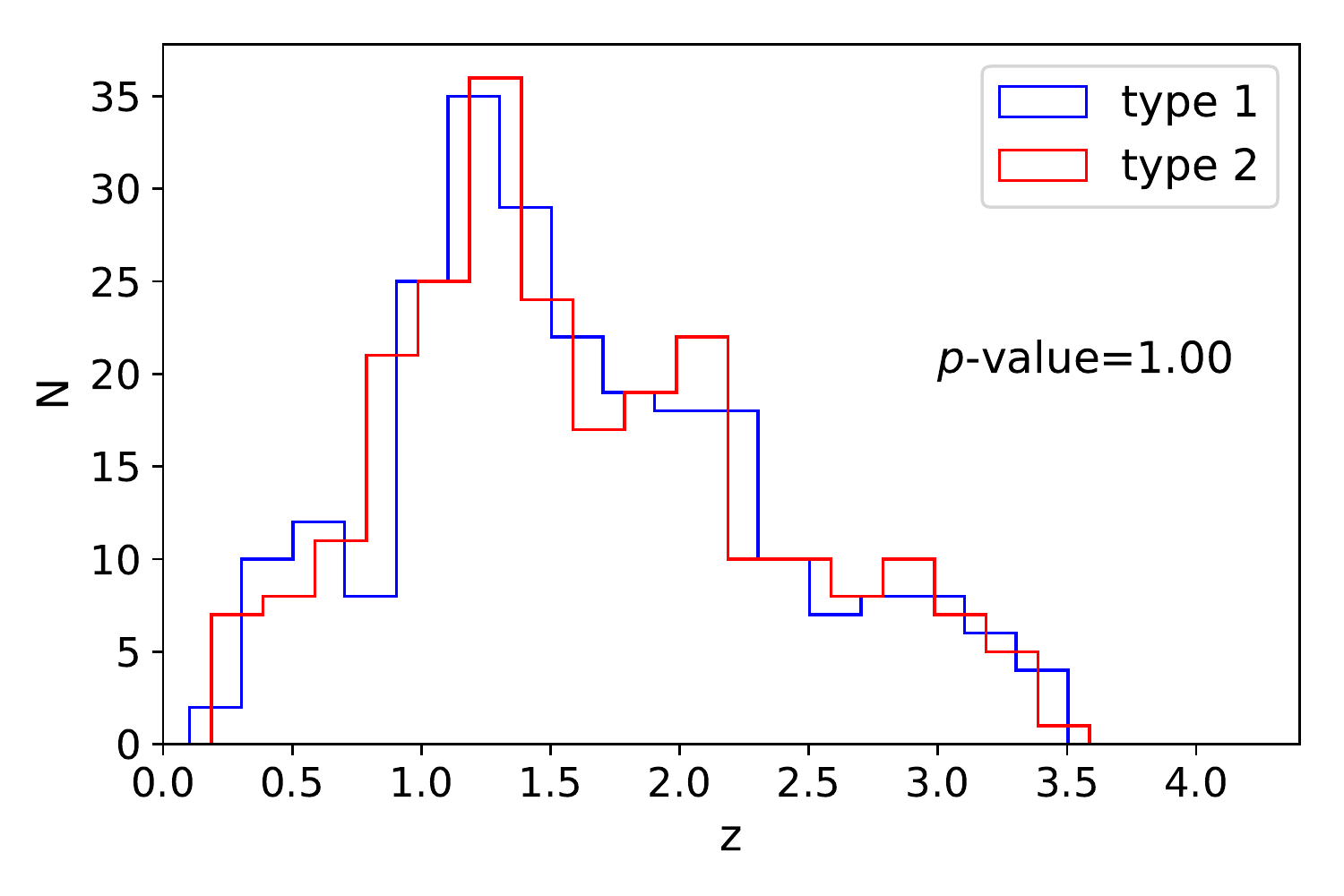}
\includegraphics{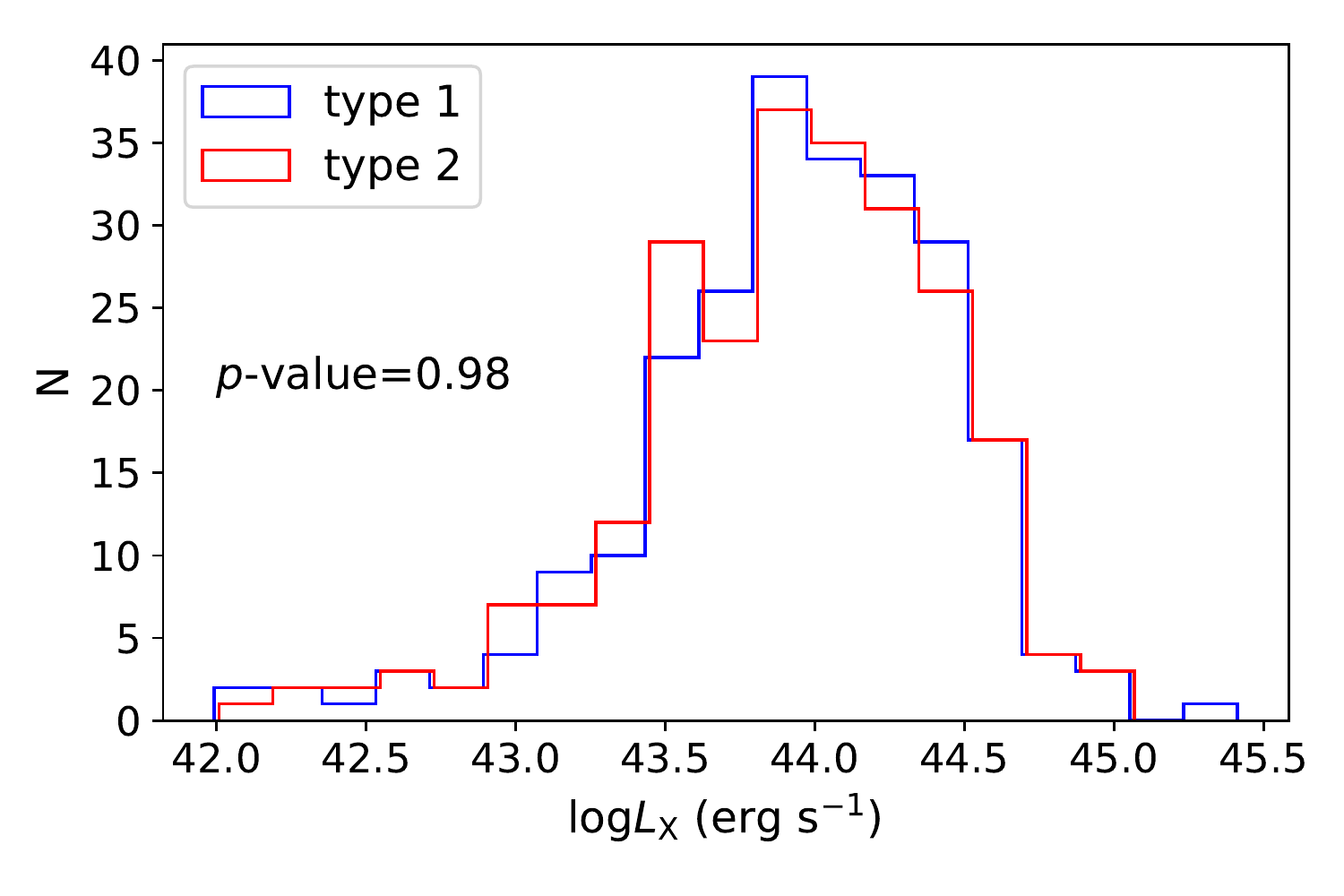}
}
\caption{Example distributions of $z$ (\emph{left}) and $L_\mathrm{X}$ (\emph{right}) for type~1 (blue) and type~2 (red) samples with controlled $z$ and $L_\mathrm{X}$ when analyzing $M_\star$ in Case~1. The distributions appear to be similar as expected.\label{controlDistFig}}
\end{figure*}
\begin{figure}[htb!]
\resizebox{\hsize}{!}{
\includegraphics{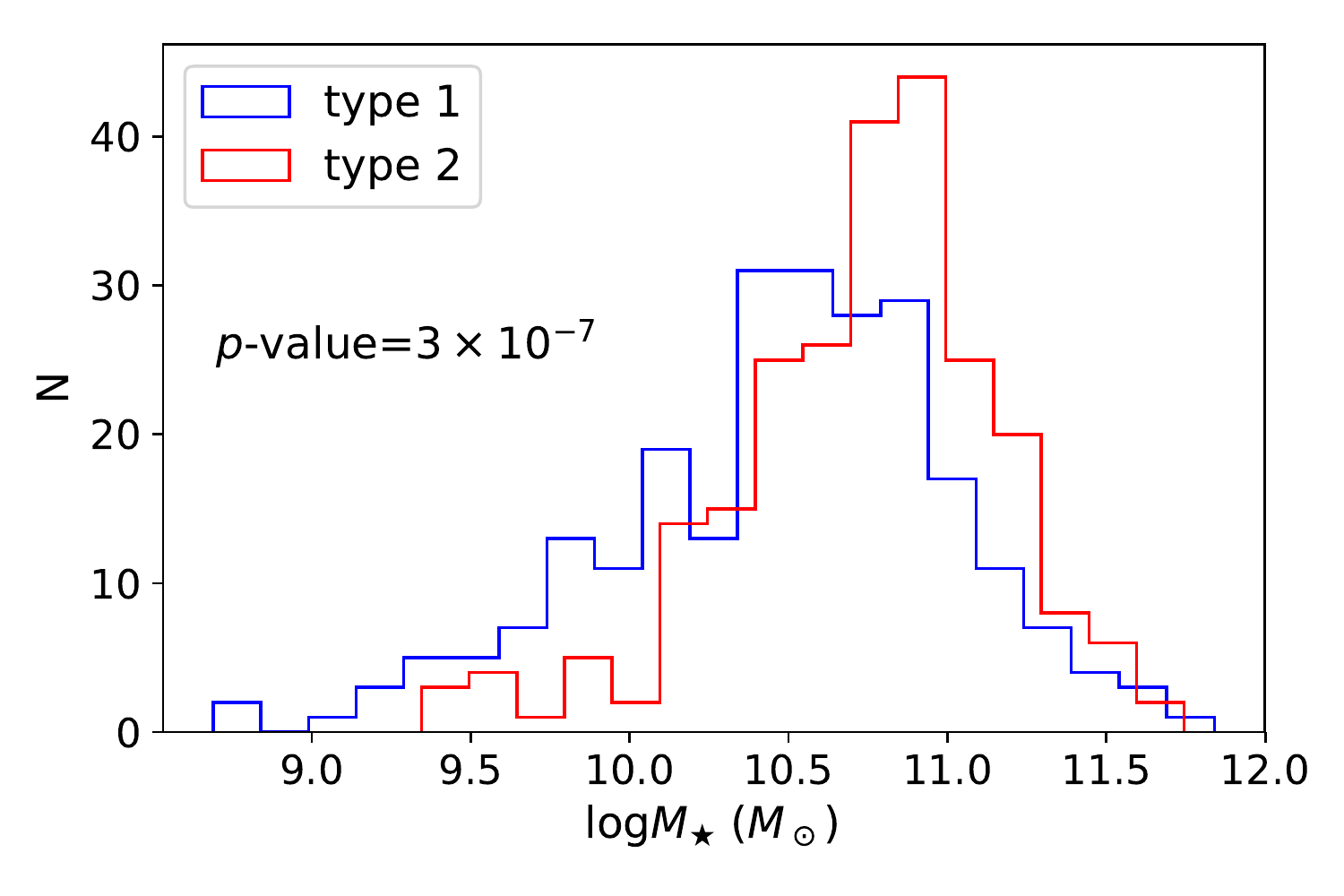}
}
\caption{Example distributions of $M_\star$ for type~1 (blue) and type~2 (red) hosts with controlled $z$ and $L_\mathrm{X}$ in Case~1.\label{MstarDistFig}}
\end{figure}

The new samples change every time we repeat the above procedures because we select sources randomly. We only show a typical example of the new samples in Case~1 here. We select 241 type~1 sources and 241 type~2 sources. Fig.~\ref{controlDistFig} displays the distributions of $z$ and $L_\mathrm{X}$ for the 482 selected sources, which visually shows the similarity of their distributions for the two different types of AGNs after controlling for $z$ and $L_\mathrm{X}$. We also use a two-sample Kolmogorov-Smirnov test (KS test) to examine their consistency: the \emph{p}-values of $z$ and $L_\mathrm{X}$ are 1.00 and 0.98, respectively, indicating that the two parameters have been controlled acceptably. Distributions of $M_\star$ are displayed in Fig.~\ref{MstarDistFig}. The figure shows that the $M_\star$ of type~1 hosts tend to be smaller than those of type~2 hosts. The KS test shows that the difference is statistically significant with \emph{p}-value = $3\times10^{-7}$ ($\approx5\sigma$), which is much smaller than our nominal \emph{p}-value 0.05 ($2\sigma$; see Section~\ref{section: intro}).\par
We utilize a bootstrapping method to calculate errors of mean log$M_\star$ ($\overline{\mathrm{log}M_\star}$) for type~1 and type~2 sources and their difference ($\Delta\overline{\mathrm{log}M_\star}=\overline{\mathrm{log}M_\star}_\mathrm{type~1}-\overline{\mathrm{log}M_\star}_\mathrm{type~2}$). First, we perform bootstrap resampling on our samples 1000 times, and repeat the above procedures, i.e., controlling for $z$ and $L_\mathrm{X}$, deriving host $\overline{\mathrm{log}M_\star}$ for each type of AGN, and calculating $\Delta\overline{\mathrm{log}M_\star}$. Then we can obtain hundreds of $\overline{\mathrm{log}M_\star}$ and $\Delta\overline{\mathrm{log}M_\star}$. Finally, we can derive their mean values and (84th--16th percentile)/2 of their distributions, and the latter values are adopted as $1\sigma$ uncertainties. It is necessary to calculate the error of $\Delta\overline{\mathrm{log}M_\star}$ straightforwardly using the bootstrapping method instead of estimating it based on the log$M_\star$ errors of type~1 and type~2 sources. The latter method to estimate the error of $\Delta\overline{\mathrm{log}M_\star}$ is incorrect because it ignores the covariance between $\overline{\mathrm{log}M_\star}_\mathrm{type~1}$ and $\overline{\mathrm{log}M_\star}_\mathrm{type~2}$. Indeed, $\overline{\mathrm{log}M_\star}_\mathrm{type~1}$ and $\overline{\mathrm{log}M_\star}_\mathrm{type~2}$ are correlated with each other after controlling for $z$ and $L_\mathrm{X}$, and thus their covariance is not zero. Table~\ref{MstarerrTable} shows $\overline{\mathrm{log}M_\star}$ and $\Delta\overline{\mathrm{log}M_\star}$ for both types of AGN hosts in Case~1 and Case~2, which indicates that type~1 sources have slightly smaller $M_\star$ with a difference around 0.2 dex. The significance that the difference deviates from 0 is around $4\sigma$.\par
As a reliability check, we explore whether the difference in $M_\star$ is sensitive to our parameter settings in SED fitting by adopting other settings in the AGN template \texttt{fritz2006} in \texttt{CIGALE}. \citet{Buat15} argued that setting a large angle between the line of sight and accretion disk $\psi$ (e.g., $80^\circ$ or $90^\circ$) to represent type~1 AGNs may lead to an unrealistic contribution from the AGN component in the UV band. Therefore, they adopt different settings for type~1 AGNs ($\psi=0^\circ$ and optical depth at $9.7~\mu\mathrm{m}$ of $\tau=1.0$) from our settings in Section~\ref{section: sedfitting} ($\psi=90^\circ$ and $\tau=6.0$). We follow their settings and label this case as Case~3. The results of $M_\star$ analyses in Case~3 are displayed in Table~\ref{MstarerrTable}. From the table, we can see that type~1 sources do have slightly smaller $M_\star$ no matter which samples and parameter settings are adopted.\par
\begin{table}[hpbt]
\centering
\caption{
$\overline{\mathrm{log}M_\star}$ of type~1 and type~2 samples.}
\label{MstarerrTable}
\begin{threeparttable}
\begin{tabular}{cccc}
\hline
\hline
Case & $\overline{\mathrm{log}M_\star}_\mathrm{type~1}$ & $\overline{\mathrm{log}M_\star}_\mathrm{type~2}$ & Difference\\
\hline
Case~1 & $10.48\pm0.04$ & $10.71\pm0.03$ & $-0.22\pm0.05$\\
Case~2 & $10.51\pm0.02$ & $10.63\pm0.02$ & $-0.12\pm0.03$\\
Case~3 & $10.52\pm0.03$ & $10.71\pm0.03$ & $-0.19\pm0.04$\\
\hline
\hline
\end{tabular}
\begin{tablenotes}
\small
\item
\emph{Notes.} Case~1 and Case~2 are defined in Section~\ref{section: classification}. We follow parameter settings of the AGN component in \citet{Buat15} in Case~3, and the samples are the same as those in Case~1. $M_\star$ of type~2 populations seem to be slightly higher than those of type~1 sources.
\end{tablenotes}
\end{threeparttable}
\end{table}

\begin{figure}[htb!]
\resizebox{\hsize}{!}{
\includegraphics{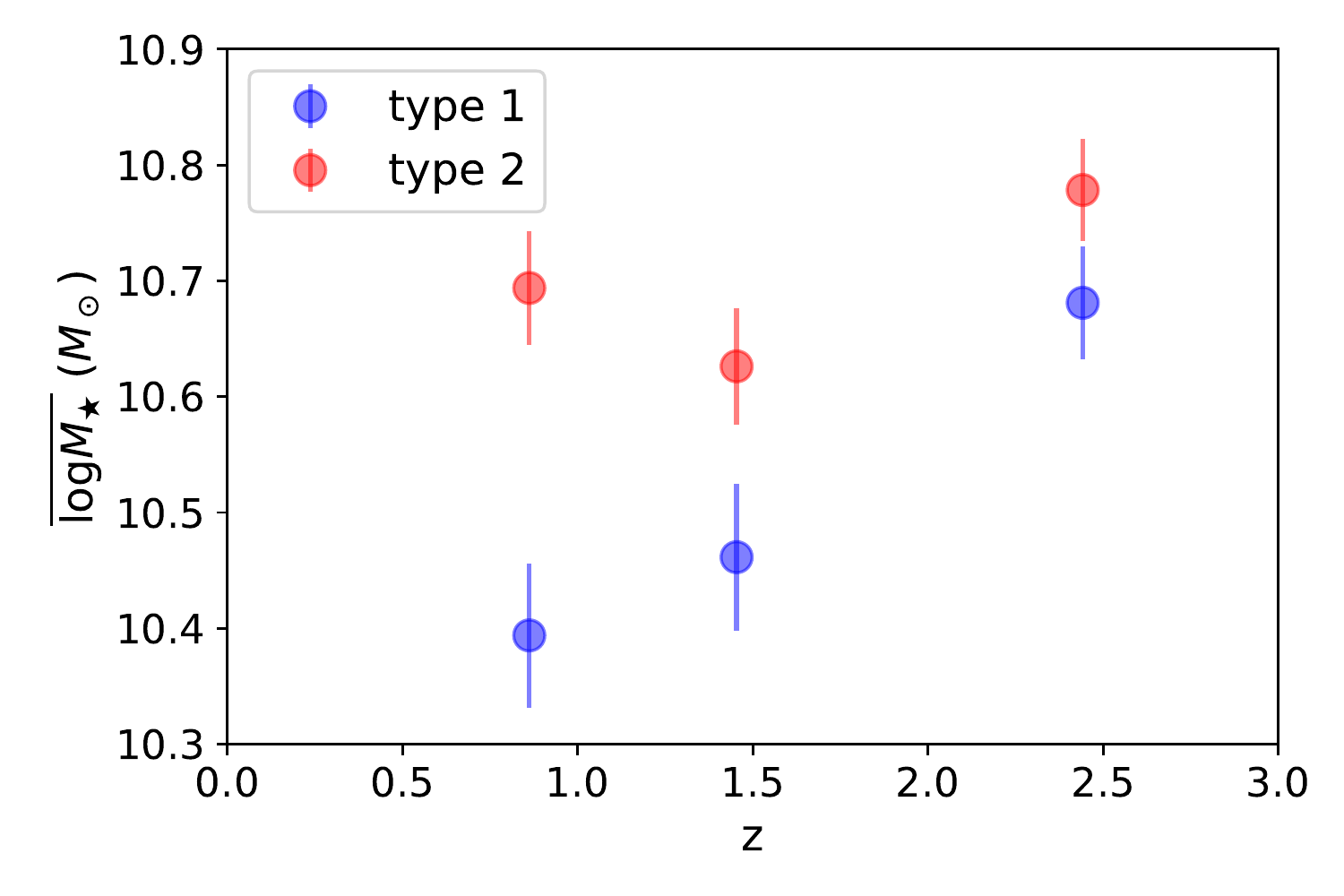}
}
\caption{
$\overline{\mathrm{log}M_\star}$ versus $z$ of Case~1 in three redshift bins: $z<1.2$, $1.2\le z<1.8$ and $z\ge1.8$. Redshift and $L_\mathrm{X}$ of the two populations are controlled to be similar. Type~1 AGNs tend to have lower host $M_\star$.
\label{MstarEvoFig}
}
\end{figure}

To probe a possible redshift evolution of such an $M_\star$ difference, we show $M_\star$ as a function of $z$ in Fig.~\ref{MstarEvoFig}. The figure indicates that $M_\star$ of type~1 host galaxies is always smaller than that of type~2 hosts in each redshift bin. Besides, there is not an apparent dependence on $z$ of $M_\star$.

\subsection{Star-Formation Rate}
\label{section: SFR}
\subsubsection{SFRs In Our Sample}
\label{section: SFRSec1}
Generally, there are two methods to calculate SFR for AGN host galaxies: one is SED fitting, as we described in Section~\ref{section: sedfitting}; the other method that is widely used is based on single-band FIR photometry. Previous studies based on different SFR measurement techniques reached qualitatively different conclusions about SFRs for type 1 versus type 2 hosts (e.g., \citealt{Merloni14, Bornancini18}). To investigate this controversy, we estimate SFRs based on both SED fitting and FIR flux techniques, respectively.\par
In the latter method, FIR SEDs are assumed to be mainly contributed by cold dust whose temperatures do not vary significantly among different sources. We follow the method in \citet{calSFR} and \citet{GuangSFR} to calculate FIR-based SFR with FIR ($100~\mu\mathrm{m}$, $160~\mu\mathrm{m}$, $250~\mu\mathrm{m}$, $350~\mu\mathrm{m}$, and $500~\mu\mathrm{m}$) fluxes. For FIR photometry at wavelength $\lambda$, we calculate the ratio between the observed flux $S_\lambda$ and the monochromatic flux of a given template at the corresponding observed-frame wavelength $\lambda$ ($S_\lambda^\mathrm{T}$), and derive the total infrared photometry $L_\mathrm{IR}$ using the following equation:
\begin{align}
L_\mathrm{IR}=\frac{S_\lambda}{S_\lambda^\mathrm{T}}L_\mathrm{IR}^\mathrm{T},\label{FIR2SFR}
\end{align}
where $L_\mathrm{IR}^\mathrm{T}$ is the $8-1000~\mu\mathrm{m}$ luminosity of the adopted template.\par
Here, we use the templates in \citet{SEDK12}, among which a ``$z\sim1$ SF galaxies'' template with $L_\mathrm{IR}^\mathrm{T}=4.26\times10^{11}L_\odot$ is adopted for sources at $z\le1.5$ and a ``$z\sim2$ SF galaxies'' template with $L_\mathrm{IR}^\mathrm{T}=2.06\times10^{12}L_\odot$ is adopted for sources at $z>1.5$. The priority of the adopted wavelength is $500~\mu\mathrm{m}>350~\mu\mathrm{m}>250~\mu\mathrm{m}>160~\mu\mathrm{m}>100~\mu\mathrm{m}$. Such a priority level is used to avoid AGN contamination as much as possible (e.g., \citealt{Stanley17}).\par
\begin{figure}[htb!]
\resizebox{\hsize}{!}{
\includegraphics{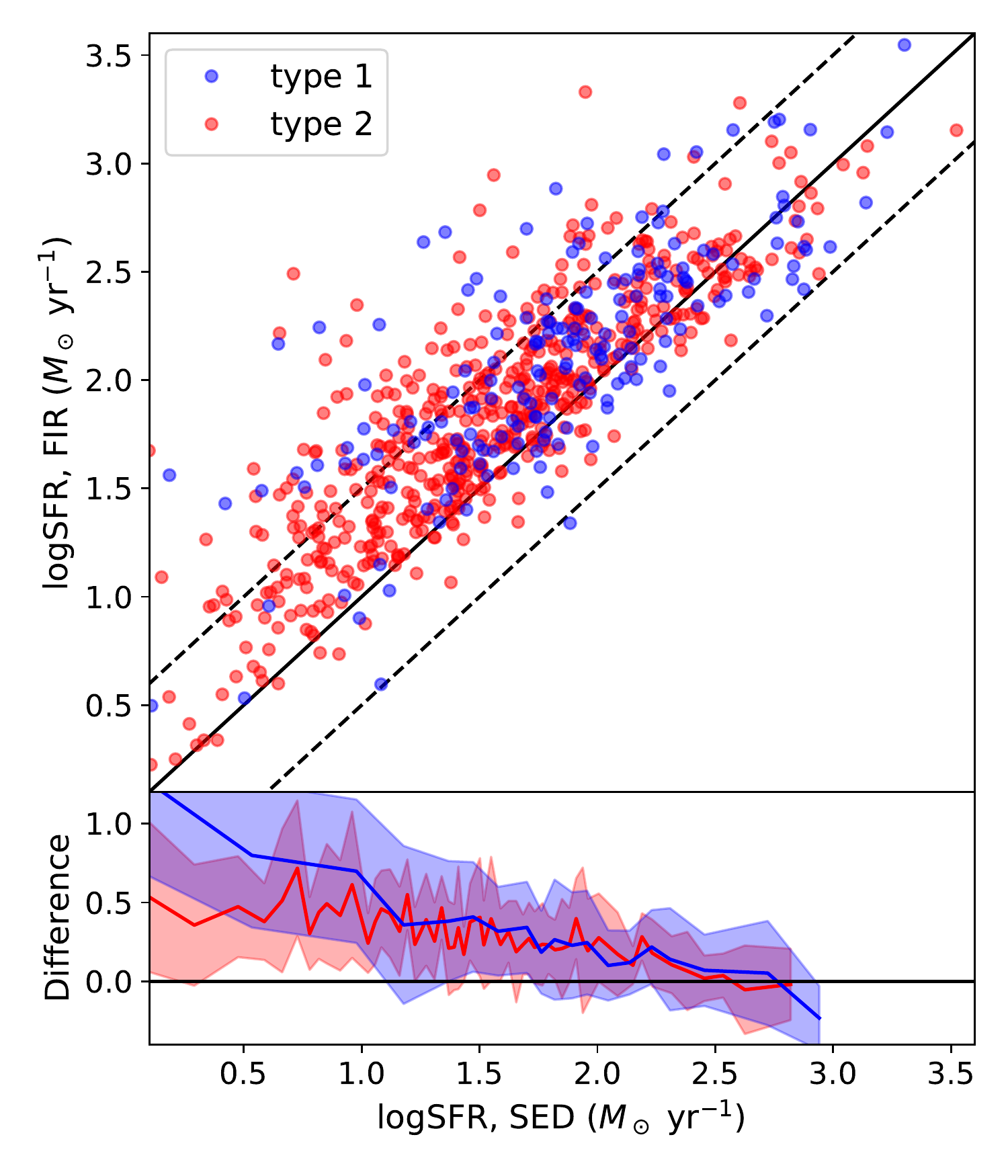}
}
\caption{
\emph{Top}: comparison between FIR-based SFRs and SED-based SFRs for sources detected by \emph{Herschel} in Case~2. The solid black line indicates a one-to-one relationship between the two kinds of SFRs. The dashed black lines indicate 0.5 dex offsets from the solid black line. \emph{Bottom}: the difference between these two kinds of SFRs versus SED-based SFRs, where the difference is defined as $\mathrm{log}[(\mathrm{SFR,~FIR})/(\mathrm{SFR,~SED})]$. Blue and red segmented lines indicate running mean SFR offsets from bins of 10 sources for type~1 and type~2 sources, respectively, and the shaded regions are standard deviations of the differences.
\label{Comp2SFR}}
\end{figure}

Table~\ref{DetectFrac} shows the fractions of sources in our sample detected in each FIR band, and only 32.1\% of our sources are detected in at least one FIR band. We can derive SFRs from FIR photometry for these detected sources. These sources also have reliable SED-based SFRs because \texttt{CIGALE} adopts energy conservation to constrain the stellar component based on the FIR photometry (Section~\ref{section: sedfitting}). Fig.~\ref{Comp2SFR} compares FIR-based SFRs with SED-based SFRs in Case~2. As the figure shows, these two kinds of SFRs are generally consistent within 0.5 dex for sources with $\mathrm{logSFR}\gtrsim0.5$, but FIR-based SFRs tend to be larger than SED-based SFRs. Such a bias is unlikely mainly caused by blending issues, i.e., the flux of a faint source beside a bright source may be systematically overestimated (e.g., \citealt{Magnelli14}). Indeed, the \emph{Herschel} observations in the COSMOS field are relatively shallow \citep{Oliver12}, and thus blending is not strong. Furthermore, the \texttt{XID+} tool that was used to extract \emph{Herschel} fluxes in COSMOS has been proved to mitigate this issue well and thus should give reliable estimates of the fluxes of crowded sources \citep{HerMES}. The bias might be primarily driven by a selection effect, as reported in \citet{Bongiorno12} and \citet{GuangSFR}: at a given SED-based SFR, \emph{Herschel}-detected sources tend to have higher FIR fluxes and thus higher FIR-based SFRs. The bias is more significant for sources with low SED-based SFRs because it is more difficult for sources with low SFRs to reach the detection threshold; thus, the detected ones are more likely to reside in the upper envelope of the $\mathrm{SFR}_\mathrm{FIR}/\mathrm{SFR}_\mathrm{SED}$ distribution. The bottom panel of Fig.~\ref{Comp2SFR} also indicates that the difference between the two SFR measurements is similar for both types of sources. Therefore, systematic errors on derived SFRs depend weakly on AGN type for sources within this SFR range, indicating that AGN light pollution for FIR-detected type~1 AGNs is unlikely to affect our SED-fitting results systematically.\par
\begin{table*}[hpbt]
\centering
\caption{Fractions of sources detected in each FIR band in our sample.}
\label{DetectFrac}
\begin{threeparttable}
\begin{tabular}{cccccc}
\hline
\hline
Wavelength ($\mu\mathrm{m}$) & 100 & 160 & 250 & 350 & 500\\
\hline
Detection rate & 15.8\% & 14.0\% & 26.6\% & 17.1\% & 5.9\%\\
\hline
\hline
\end{tabular}
\begin{tablenotes}
\small
\item
\emph{Notes.} For the same observing instrument (PACS at $100~\mu\mathrm{m}$ and $160~\mu\mathrm{m}$; SPIRE at $250~\mu\mathrm{m}$, $350~\mu\mathrm{m}$, and $500~\mu\mathrm{m}$), the detection rate decreases as wavelength increases.
\end{tablenotes}
\end{threeparttable}
\end{table*}

\begin{figure*}[htb!]
\resizebox{\hsize}{!}{
\includegraphics{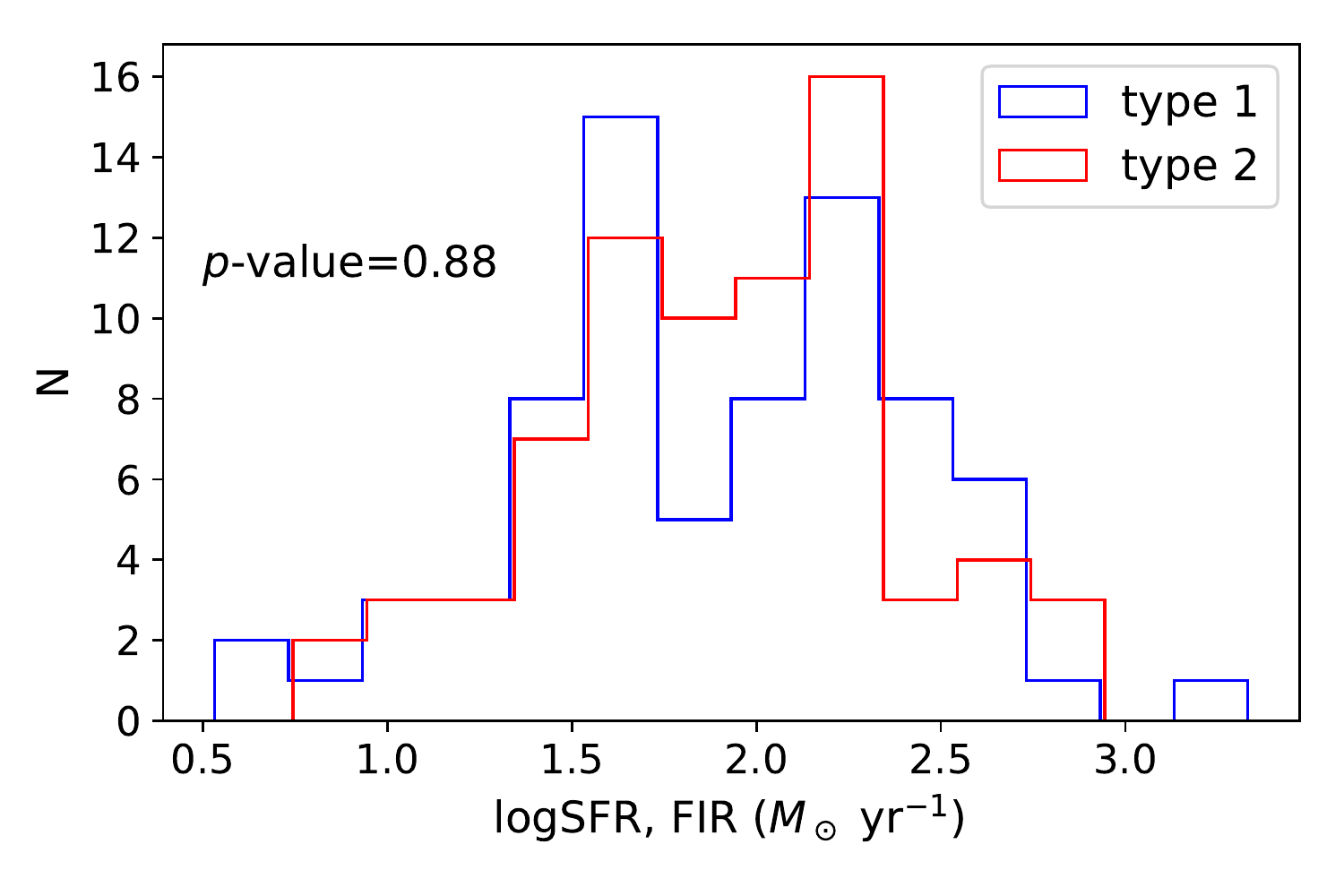}
\includegraphics{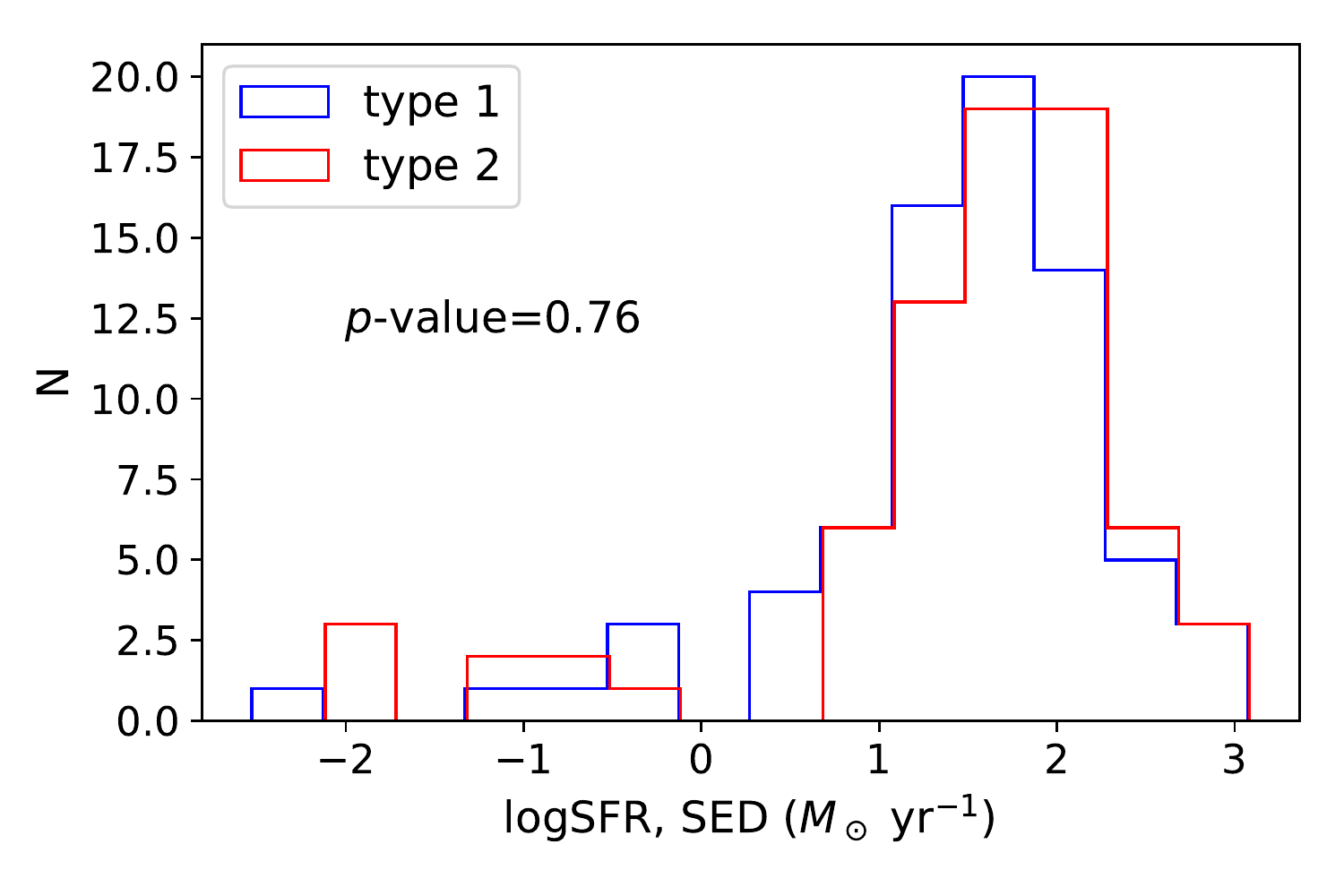}
}
\caption{
Example distributions of FIR-based SFRs (\emph{left}) and SED-based SFRs (\emph{right}) for \emph{Herschel}-detected type~1 (blue) and type~2 (red) sources in Case~1. Their $z$ and $L_\mathrm{X}$ are controlled to be similar. This figure shows that FIR-detected populations have similar SFRs.
\label{SFRDist_FIR}
}
\end{figure*}

As the first step to probe possible differences in SFRs between the two types of AGNs, we compare their host SFRs for FIR-detected sources. We control for $z$ and $L_\mathrm{X}$ simultaneously using the same method as that in Section~\ref{section: Mstar}. As a typical example, we select 74 sources of each type in Case~1 after controlling for the parameters. We plot their FIR-based and SED-based SFRs in Fig.~\ref{SFRDist_FIR}. Type~1 and type~2 sources that are detected in the FIR band appear to have similar FIR-based SFRs with \emph{p}-value = 0.88 using the KS test, as well as SED-based SFRs with \emph{p}-value = 0.76. We also repeat the procedure in several redshift bins and do not find statistically significant differences in SFRs for FIR-detected type~1 and type~2 populations. Therefore, we conclude that FIR-detected type~1 and type~2 sources have similar SFRs, as shown by both FIR-based and SED-based SFR measurements.\par
For sources undetected by \emph{Herschel}, SFRs derived from SED fitting may be affected by contamination from AGN emission at UV/optical wavelengths. Therefore, we stack \emph{Herschel} fluxes to calculate typical FIR-based SFRs. Since sources are highly crowded in the SPIRE maps, we do not stack undetected sources in the SPIRE maps. For the PACS maps, we only stack sources in the $160~\mu\mathrm{m}$ map that is less affected by AGN contamination than the $100~\mu\mathrm{m}$ map. We also find that all the coverages of our sources are larger than half of the central coverage of the map. Therefore, our stacking procedure is not significantly affected by considerable noise in low-coverage regions which may reduce the reliability of stacking \citep{Santini14Stack}.\par
To stack fluxes, we calculate net flux in the $160~\mu\mathrm{m}$ residual map for each FIR-undetected source and adopt fluxes in the PEP catalog for FIR-detected sources. We derive mean values of the fluxes as stacked fluxes for type~1 and type~2 populations in three redshift bins: $z<1.4$, $1.4\le z<2.0$, and $z\ge2.0$. Then we convert the stacked fluxes to stacked average SFRs using Equation \ref{FIR2SFR} with the average redshifts of the sources in each redshift bin. Finally, we average the SFRs in the three redshift bins weighted by the number of selected sources in each bin, and the resulting value is adopted as mean FIR-based SFR in the whole redshift range.\par
We show mean FIR-based SFRs and mean SED-based SFRs of both populations, as well as their differences after controlling for $z$ and $L_\mathrm{X}$ in Case~1 and Case~2 in Table~\ref{StackTable}. The differences are defined as the values of the type~1 sample subtracting those of the type~2 sample. Again, we use the bootstrapping method described in Section~\ref{section: Mstar} to estimate the errors. The results of both cases indicate that type~1 and type~2 sources generally have similar average SFRs, while the FIR-based SFRs in Case~2 seem to be different. The significance of the difference is $2.04\sigma$, which is slightly more significant than our threshold ($2\sigma$; see Section~\ref{section: intro}). Since we have four trials considered in Table~\ref{StackTable}, it is not surprising to find a $2.04\sigma$ deviation for a single trial (cf., the Bonferroni correction). This $2.04\sigma$ deviation corresponds to an overall significance of only $1.4\sigma$ for the whole sample based on the Bonferroni correction, and thus the deviation may simply be due to statistical fluctuations. Therefore, the host SFRs of type~1 and type~2 AGNs appear similar. We note that this similarity in average SFR holds not only for the FIR-based SFR but also for the SED-based SFR, indicating that the SED-based SFR measurements are not significantly biased. We attribute this reliability of the SED-based SFRs to the inclusion of FIR photometry (and upper limits) in the SED fitting, which effectively constrains the stellar population based on energy conservation (Section~\ref{section: sedfitting}).\par
\begin{figure}[htb!]
\resizebox{\hsize}{!}{
\includegraphics{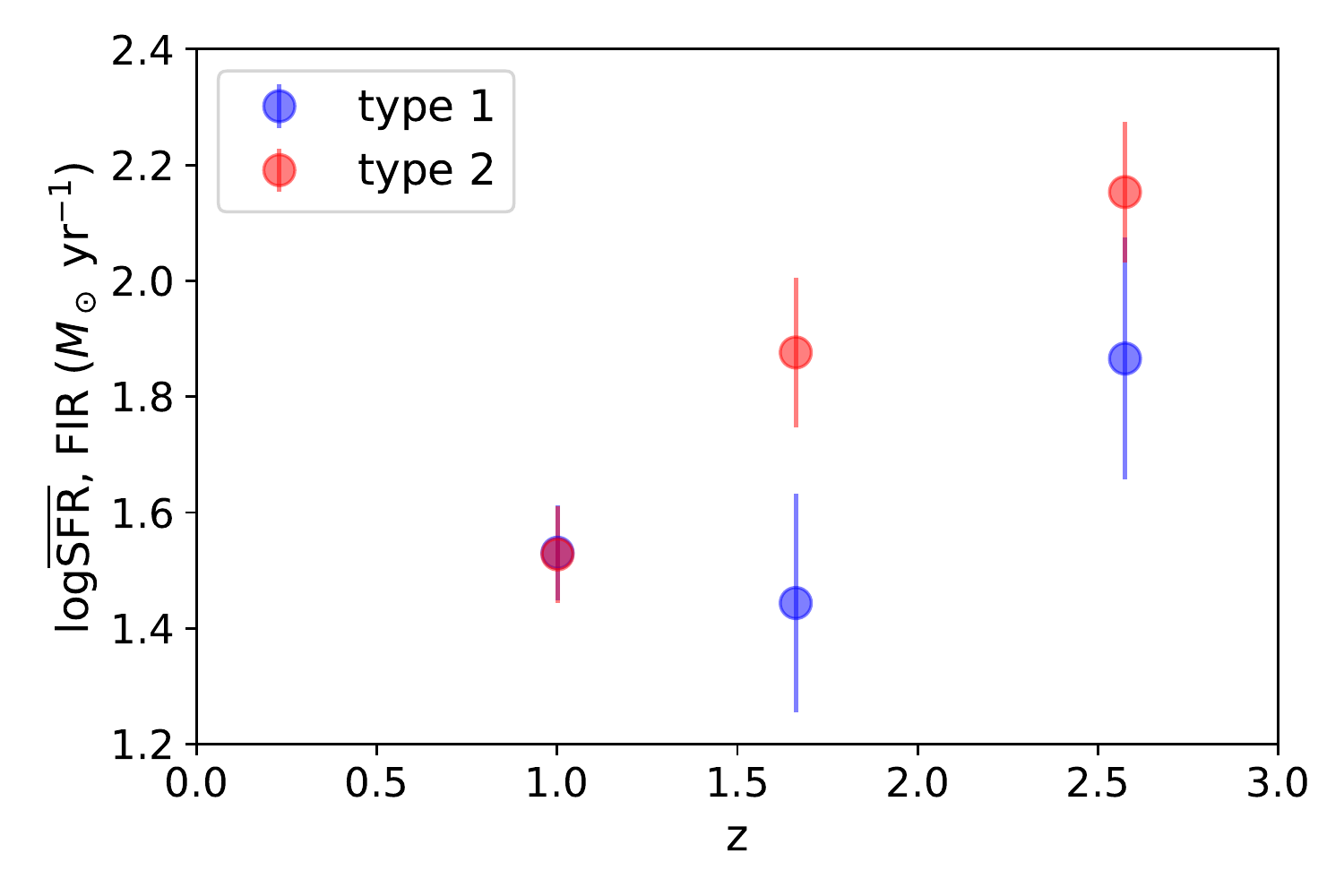}
}
\caption{Average SFRs of type~1 (blue) and type~2 (red) sources in Case~1 in each redshift bin ($z<1.4$, $1.4\le z<2.0$ and $z\ge2.0$). Redshift and $L_\mathrm{X}$ of the two populations are controlled to be similar. The SFRs are similar for type~1 and type~2 sources.\label{z_SFRFig}}
\end{figure}

We also calculate FIR-based average SFRs in Case~1 in three redshift ranges: $z<1.4$, $1.4\le z<2.0$, and $z\ge2.0$, and the results are displayed in Fig.~\ref{z_SFRFig}. Note that the significance of the difference at $1.4\le z<2.0$ is only $1.9\sigma$, below our nominal significance threshold ($2\sigma$). Therefore, the figure indicates that type~1 and type~2 AGNs have similar average SFRs over the whole redshift range, and there is no apparent redshift trend of the difference in the SFRs.
\begin{table*}[hpbt]
\caption{Average SFRs of type~1 and type~2 samples.}
\label{StackTable}
\centering
\begin{threeparttable}
\begin{tabular}{c|ccc|ccc}
\hline
\hline
\multirow{2}{*}{Parameter} & \multicolumn{3}{c|}{Case~1} & \multicolumn{3}{c}{Case~2}\\
\cline{2-7}
& Type~1 & Type~2 & Difference & Type~1 & Type~2 & Difference\\
\hline
$\mathrm{log}\overline{\mathrm{SFR}}$, FIR ($M_\odot~\mathrm{yr}^{-1}$) & $1.62\pm0.11$ & $1.87\pm0.08$ & $-0.25\pm0.13$ & $1.72\pm0.07$ & $1.89\pm0.05$ & $-0.17\pm0.08$\\
$\mathrm{log}\overline{\mathrm{SFR}}$, SED ($M_\odot~\mathrm{yr}^{-1}$) & $1.98\pm0.07$ & $1.88\pm0.05$ & $0.11\pm0.07$ & $2.06\pm0.10$ & $1.90\pm0.04$ & $0.16\pm0.11$\\
\hline
\hline
\end{tabular}
\begin{tablenotes}
\small
\item
\emph{Notes.} The ``$\mathrm{log}\overline{\mathrm{SFR}}$, FIR'' is converted from stacked $160~\mu\mathrm{m}$ fluxes. Redshift and $L_\mathrm{X}$ are controlled to be similar for type~1 and type~2 sources. Both cases are defined in Section~\ref{section: classification}. The differences are defined as the values of the type~1 sample subtracting those of the type~2 sample.
\end{tablenotes}
\end{threeparttable}
\end{table*}

\subsubsection{A Further Test For The SFR Measurements}
We note that both methods of deriving SFR mentioned above rely on FIR photometry more or less, and thus they are not exactly independent. Therefore, we use the $H_\alpha$ luminosity as an independent SFR indicator and perform a simple cross-test for the methods among inactive galaxies.\par
The observed $H_\alpha$ fluxes are from the Fiber Multi-Object Spectrograph (FMOS)-COSMOS Survey \citep{Silverman15, Kashino18}. We select 1472 sources based on the following criteria: they have available $H_\alpha$ fluxes in the FMOS-COSMOS Survey so that we can derive their $H_\alpha$-based SFRs; they are included in the COSMOS2015 catalog so that we can conduct SED fitting for them to obtain their SED-based SFRs; they are undetected in the COSMOS-Legacy Survey so that they are normal galaxies instead of active galaxies generally. The last requirement aims to prevent AGNs from contaminating the observed $H_\alpha$ fluxes, and thus the $H_\alpha$ fluxes should be reliable. To obtain the intrinsic fluxes for these sources, we correct the observed fluxes based on the Calzetti extinction law \citep{dustatt_calzleit}:
\begin{align}
F_i(H_\alpha)&=F_o(H_\alpha)10^{0.4E_n(B-V)k(H_\alpha)},~\mathrm{and}\\
k(H_\alpha)&=3.325,
\end{align}
where $F_i(H_\alpha)$ and $F_o(H_\alpha)$ are the intrinsic and observed $H_\alpha$ fluxes, respectively; $E_n(B-V)$ is the color excess for nebular gas emission lines, which is derived from SED fitting. We then use the calibration in Section 2.3 of \citet{Kennicutt98} to convert the $H_\alpha$ luminosities $L(H_\alpha)$ to SFRs:
\begin{align}
\mathrm{SFR}(M_\odot~\mathrm{yr}^{-1})=7.9\times10^{-42}L(H_\alpha)(\mathrm{erg~s}^{-1})
\end{align}\par
\begin{figure}[htb!]
\resizebox{\hsize}{!}{
\includegraphics{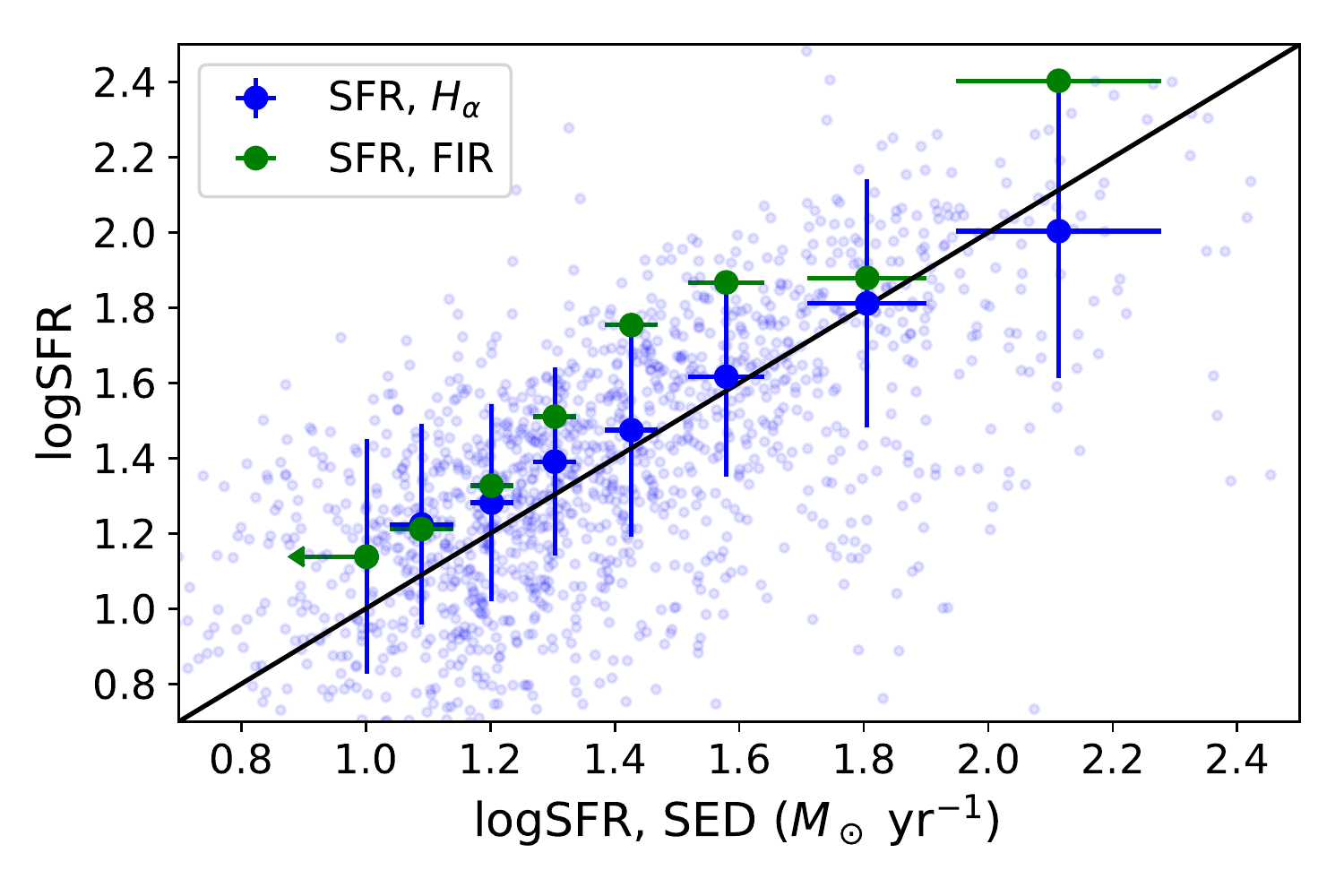}
}
\caption{Comparison between $H_\alpha$-based (blue) or FIR-based (green) SFRs and SED-based SFRs. The small semitransparent blue points are $H_\alpha$-based SFRs for each source. The large opaque points with error bars are stacked or average SFRs in different abscissa bins. The errors are estimated from ($\mathrm{84th–16th}$ percentile)/2 of the SFR distributions in each bin. The black line is a one-to-one relationship. This figure shows that these three SFR measurements are generally consistent with each other.\label{SFRHalphaFig}}
\end{figure}

We compare $H_\alpha$-based SFRs, FIR-based SFRs, and SED-based SFRs in Fig. \ref{SFRHalphaFig}. All the sources have $H_\alpha$-based and SED-based SFRs, but some lack FIR-based SFRs. Therefore, we stack the FIR photometry to obtain the average FIR-based SFRs, and the stacking method has been described in Section \ref{section: SFRSec1}. The figure shows that the three SFR measurements are generally consistent with each other.\par
This additional independent cross-check further justifies the reliability of our SFR measurements. However, it is infeasible to further quantitatively assess the sensitivity of our methods in detecting the SFR differences of our AGN samples. This is because, to do this, we need $H_\alpha$ samples that have sample sizes, redshift distributions, and $M_\star$ distributions matched with the AGN samples. However, the $H_\alpha$ sample size of FMOS-COSMOS is not sufficiently large to perform such a task. Also, non-negligible measurement uncertainties of $H_\alpha$-based SFRs likely exist, and the $H_\alpha$ method might not be sufficiently precise to gauge the accuracy of the FIR and SED methods.

\subsection{Cosmic Environment}
\label{section: Environment}
We match our sources with the catalog containing environment measurements in \citet{Yang18Env} with a matching radius of $0.5''$ and obtain 1996 matched sources. The unmatched sources are mainly near the edge of the field and/or large masked regions in the COSMOS2015 survey, and they are filtered out in \citet{Yang18Env} to improve the reliability of density measurements.\par
We use $\mathrm{log}(1+\delta)$ as the indicator for overdensity (see Section~\ref{section: DefEnv} for its definition). We control for $z$ and $M_\star$ using the method in Section~\ref{section: Mstar}, and show the results in Case~1 and Case~2 in Table~\ref{funcdetTable}. The results in both cases indicate that the overdensities are similar for the two AGN populations. Therefore, AGN type is unrelated to the local environment, at least in the case of surface number density.\par

\begin{table*}[hpbt]
\caption{$\overline{\mathrm{log}(1+\delta)}$ of type~1 and type~2 samples.}
\label{funcdetTable}
\centering
\begin{threeparttable}
\begin{tabular}{c|ccc|ccc}
\hline
\hline
\multirow{2}{*}{Redshift range} & \multicolumn{3}{c|}{Case~1} & \multicolumn{3}{c}{Case~2}\\
\cline{2-7}
& Type~1 & Type~2 & Difference & Type~1 & Type~2 & Difference\\
\hline
Unlimited & $0.055\pm0.017$ & $0.094\pm0.018$ & $-0.039\pm0.025$ & $0.033\pm0.011$ & $0.063\pm0.011$ & $-0.030\pm0.016$\\
$z>1.2$ & $0.023\pm0.017$ & $0.074\pm0.017$ & $-0.050\pm0.025$ & $0.020\pm0.012$ & $0.050\pm0.012$ & $-0.030\pm0.018$\\
$z<1.2$ & $0.120\pm0.031$ & $0.128\pm0.035$ & $-0.008\pm0.045$ & $0.087\pm0.025$ & $0.092\pm0.025$ & $-0.004\pm0.034$\\
\hline
\hline
\end{tabular}
\begin{tablenotes}
\small
\item
\emph{Notes.} Redshift and host $M_\star$ are controlled to be similar for both types of AGNs. Both cases are defined in Section~\ref{section: classification}. The differences are defined as $\mathrm{log}(1+\delta)_\mathrm{type~1}-\mathrm{log}(1+\delta)_\mathrm{type~2}$.
\end{tablenotes}
\end{threeparttable}
\end{table*}

On larger scales, we investigate whether a particular type of AGN tends to reside in a certain kind of cosmic web environment. As mentioned in Section~\ref{section: DefEnv}, the global environment is classified as ``field'', ``filament'', or ``cluster'', and only ``field'' and ``filament'' are defined above $z=1.2$. We examine whether the fractions of sources in field ($f_\mathrm{field}$), filament ($f_\mathrm{filament}$), and cluster ($f_\mathrm{cluster}$) environments are different for different AGN types. Again, we control for $z$ and $M_\star$ and show the results in Table~\ref{fracWebTable}. The results show that the differences in $f_\mathrm{field}$, $f_\mathrm{filament}$, and $f_\mathrm{cluster}$ between type~1 sources and type~2 sources are insignificant.\par
We are unable to divide our sources into more redshift bins because our sample size is not sufficiently large. When controlling for SFR or $L_\mathrm{X}$ as well, we also find that both the local and global environments are similar.\par

\begin{table*}[hpbt]
\caption{Fraction of sources in each kind of cosmic-web environment.}
\label{fracWebTable}
\centering
\begin{threeparttable}
\begin{tabular}{cc|ccc|ccc}
\hline
\hline
\multirow{2}{*}{Redshift range} & \multirow{2}{*}{Fraction} & \multicolumn{3}{c|}{Case~1} & \multicolumn{3}{c}{Case~2}\\
\cline{3-8}
&&  Type~1 & Type~2 & Difference & Type~1 & Type~2 & Difference\\
\hline
Unlimited & $f_\mathrm{field}$ & $0.45\pm0.04$ & $0.36\pm0.04$ & $0.09\pm0.06$ & $0.45\pm0.02$ & $0.39\pm0.03$ & $0.06\pm0.04$\\
\hline
$z>1.2$ & $f_\mathrm{field}$ & $0.52\pm0.05$ & $0.39\pm0.05$ & $0.13\pm0.07$ & $0.48\pm0.03$ & $0.41\pm0.03$ & $0.07\pm0.04$\\
\hline
\multirow{3}{*}{$z<1.2$} & $f_\mathrm{field}$ & $0.31\pm0.07$ & $0.30\pm0.07$ & $0.00\pm0.09$ & $0.33\pm0.05$ & $0.36\pm0.05$ & $-0.03\pm0.07$\\
& $f_\mathrm{filament}$ & $0.67\pm0.07$ & $0.61\pm0.07$ & $0.06\pm0.10$ & $0.63\pm0.05$ & $0.55\pm0.05$ & $0.08\pm0.07$\\
& $f_\mathrm{cluster}$ & $0.02\pm0.02$ & $0.08\pm0.04$ & $-0.06\pm0.04$ & $0.04\pm0.02$ & $0.09\pm0.03$ & $-0.05\pm0.04$\\
\hline
\hline
\end{tabular}
\begin{tablenotes}
\small
\item
\emph{Notes.} Redshift and host $M_\star$ are controlled to be similar for both types of AGNs. The ``Cluster'' environment is only defined at $z<1.2$, and thus we only show $f_\mathrm{field}$ in the whole redshift range. At $z>1.2$, $f_\mathrm{filament}$ can be straightforwardly calculated using $f_\mathrm{filament}=1-f_\mathrm{field}$ and $f_\mathrm{filament,~err}=f_\mathrm{field,~err}$. Both cases are defined in Section~\ref{section: classification}. The differences are defined as the values of the type~1 sample subtracting those of the type~2 sample.
\end{tablenotes}
\end{threeparttable}
\end{table*}

\section{Summary and Discussion}
\label{section: summary}
In this work, we use \mbox{X-ray} selected AGNs in the COSMOS field to examine whether AGN host-galaxy properties including $M_\star$, SFR, and environment are related to their AGN optical spectral types. Our main conclusions with discussion are the following:
\begin{itemize}
\item[1.]{
Type~1 AGNs have slightly smaller $M_\star$ (by about 0.2 dex) than type~2 AGNs even when $z$ and $L_\mathrm{X}$ are controlled (Section~\ref{section: Mstar}), and the difference is statistically significant ($\approx4\sigma$). We test various sample-selection criteria and parameter settings in \texttt{CIGALE}, and find that type~1 sources always have slightly lower $M_\star$. Those trials support the reliability of the difference. This difference in $M_\star$ for type~1 vs. type~2 AGNs indicates that spectral-type transformation (e.g., \citealt{Tohline76, Penston84}) is unlikely to be prevalent, because frequent wide-spread transitions would average out any differences in host-galaxy properties (see Section~\ref{section: intro}).\par
One possible explanation for the small $M_\star$ difference is based on the idea that the obscuration of AGNs is partly caused by galaxy-scale gas and dust (e.g., \citealt{Matt00, Buchner17_2}). Type~1 AGNs often have lower \mbox{X-ray} derived column densities, $N_\mathrm{H}$, than type~2 sources. If the lower $N_\mathrm{H}$ of type~1 AGNs is partly caused by lower $N_\mathrm{H}$ of galaxy-scale gas in their hosts, their $M_\star$ may also be smaller because $M_\star$ follows a positive correlation with $N_\mathrm{H}$ of galaxy-scale gas ($N_\mathrm{H}\propto M_\star^{1/3}$; \citealt{Buchner17_1}). Indeed, \citet{Lanzuisi17} also found a positive correlation between $M_\star$ and \mbox{X-ray} $N_\mathrm{H}$ for AGN. Similarly, the optical obscuration of type~2 AGNs may also be partly attributed to galaxy-scale dust (e.g., \citealt{Malkan98}), and thus type~2 hosts may tend to be more massive since massive galaxies contain more dust (e.g., \citealt{Katherine17}). Here, we note that \mbox{X-ray} and optical obscuration are not identical in terms of physical origins: the former and the latter mainly depend on the amounts of gas and dust, respectively.\par
We reiterate that the difference in $M_\star$ is quantitatively small ($\Delta\mathrm{log}M_\star\approx0.2~\mathrm{dex}$), and thus it can be ignored to first approximation. Indeed, studies often discard type~1 AGNs in coevolution studies assuming both types of AGNs have similar host $M_\star$ (e.g., \citealt{GuangSFR, Yang19}). Additionally, such a small difference may explain why some previous studies (e.g., \citealt{Merloni14, Bornancini18}) did not report a difference in $M_\star$ for the two AGN populations. Indeed, as shown in \citet{Bornancini18}, the mean $M_\star$ of type~2 sources is 0.2 dex higher than that of type~1 sources, but they ignored the difference and claimed that the $M_\star$ is similar. \citet{Merloni14}, which used a smaller sample with 1310 AGNs selected in the \emph{XMM-COSMOS} field, also found no difference in $M_\star$ for the two AGN populations. They showed that the fraction of optically obscured AGNs did not depend on $M_\star$ by dividing their sample into 20 bins in the $M_\star-L_\mathrm{X}$ plane. However, dividing into many bins reduces the significance of the difference in $M_\star$ in each bin, which may explain their conclusion that the $M_\star$ is similar.
}
\item[2.]{
We control for $z$ and $L_\mathrm{X}$ and find that the SFRs of both populations are similar (Section~\ref{section: SFR}). FIR-detected type~1 and type~2 sources have similar FIR-based SFRs and SED-based SFRs. We also stack $160~\mu\mathrm{m}$ fluxes so that we can measure mean FIR fluxes even if some sources are not detected by \emph{Herschel}, and then derive their typical SFRs from the stacked fluxes. We find that the typical SFRs are similar for type~1 and type~2 sources, and the $2\sigma$ ($3\sigma$) upper limit of the difference in mean SFR is $\approx0.3~(0.4)$ dex. Therefore, type~1 and type~2 host galaxies have similar SFRs.\par
This finding indicates that galaxy-wide star formation seems not to be connected strongly with nuclear absorption. This is consistent with other works based on the \emph{XMM}-COSMOS field (e.g., \citealt{Merloni14}) and wider \emph{Herschel} fields (e.g., \citealt{Rosario12}), which showed that SFRs are not dependent on optical or \mbox{X-ray} obscuration (e.g., \citealt{Rovilos12}). However, \citet{Bornancini18} obtained the opposite conclusion that type~1 AGN host galaxies are more likely to be star-forming ones based on the COSMOS-Legacy Survey. They used UV-optical colors as indicators of star-forming states without using FIR photometry. However, the AGN contamination in the UV-to-optical band may lead to large biases for type~1 AGN hosts, and thus may affect their judgments of the star-forming states. Some works based on luminous QSOs have also found that star formation is enhanced in X-ray absorbed sources (e.g., \citealt{Page04, Stevens05}). However, their samples differ from our samples essentially in terms of AGN luminosity and obscuration type (X-ray vs. optical), and thus their results are not directly comparable to ours.\par
Our work is based on the latest photometric data (especially \emph{Hershcel}) and \mbox{X-ray} survey data, and we utilize a larger sample and more rigorous statistical approaches to investigate the SFRs compared to previous works. Besides checking the mean FIR-based SFRs as in \citet{Merloni14}, we also examine the SED-based SFRs and the SFR distributions. Therefore, we can attest the similarity of SFRs for the two AGN host populations in a more reliable way.\par
We note that some studies proposed that AGN obscuration might be caused by nuclear starburst disks in some cases (e.g., \citealt{Ballantyne08, Hickox18}). Further studies can estimate nuclear SFR in the distant universe with the upcoming JWST mission and investigate whether type~1 and type~2 hosts have similar nuclear SFRs.
}
\item[3.]{
With respect to environment, both types of AGNs have similar local (on sub-Mpc scales) and global (on $1-10$ Mpc scales) environments after controlling for $z$ and $M_\star$ (Section~\ref{section: Environment}). The $2\sigma$ ($3\sigma$) upper limits of the differences in mean $\mathrm{log}(1+\delta)$ and $f_\mathrm{field}$ are $\approx0.05~(0.07)$ and 11\% (16\%), respectively.\par
We emphasize that the scale of the so-called ``local'' environment here is still quite large, and its typical scale is 0.5 Mpc. \citet{Jiang16} showed that the difference of environments for different types of AGNs is significant only at $\lesssim0.1~\mathrm{Mpc}$, and their environments at $\sim0.5~\mathrm{Mpc}$ are similar. Therefore, our finding on the local scale does not necessarily imply that environments on a small scale ($\lesssim0.1~\mathrm{Mpc}$) for the two populations are similar; hence galaxy interactions within dark-matter halos may also contribute to the discrimination of the two populations though we are unable to detect that. Indeed, previous works have found that type~2 AGNs have more compact environments within the small scale (e.g., \citealt{Laurikainen95, DultzinHacyan99, Koulouridis06}). Future work can probe $\lesssim0.1~\mathrm{Mpc}$ scales by comparing the incidence of galaxy pairs for type~1 and type~2 hosts in COSMOS. This should be achievable given the fact that spectroscopic and high-quality photometric redshifts are available for COSMOS (e.g., \citealt{Mundy07}).\par
On a large scale ($\gtrsim1~\mathrm{Mpc}$), our conclusion that type~1 and type~2 AGN host galaxies do not have significantly different environments is consistent with other works (e.g., \citealt{Ebrero09, Gilli09, Geach13, Jiang16}). Some works also reported tentative evidence suggesting that the host galaxies of obscured AGNs reside in denser environments (e.g., \citealt{Hickox11, DiPompeo14, Donoso14, Powell18}), while some also argued that type~1 hosts are more clustered (e.g., \citealt{Allevato11, Allevato14}). However, their statistical significances are only $2-4\sigma$ generally, and their criteria to separate ``obscured'' and ``unobscured'' sources are different from ours sometimes -- they select obscured AGNs in infrared or \mbox{X-ray} bands. Therefore, our conclusion is not contradictory to theirs.\par
Our conclusion on the environment indicates that the nuclear obscuration of the AGN is not strongly linked to its environment. Indeed, \citet{Yang18Env} found that the average SMBH accretion rate generally did not depend on cosmic environment once $M_\star$ was controlled, indicating that large-scale environment might not play an important role in the evolution of SMBHs.\par
However, it is still possible that the uncertainties prevent us from finding an existing subtle difference of the environments, and we need a larger sample to study the environment better. The ongoing $12~\mathrm{deg}^2$ XMM-Spitzer Extragalactic Representative Volume Survey (XMM-SERVS) will provide suitable detections of \mbox{$\approx12000$} AGNs in well-characterized multi-wavelength fields (e.g., \citealt{Chen18}), and the next-generation all-sky \mbox{X-ray} survey conducted by \emph{eROSITA} is expected to detect millions of AGNs (e.g., \citealt{Kolodzig13_2, Kolodzig13_1}). Thus they are likely to extend our knowledge on this topic.
}
\item{
Overall, our findings show that the unified model is not strictly correct since type~1 AGNs have smaller host $M_\star$ than type~2 AGNs. The difference in $M_\star$ indicates that both host galaxy and torus may contribute to the optical obscuration of AGNs. We do not find strong evidence supporting the merger-driven coevolution model (Section \ref{section: intro}). Future simulations of the SMBH-galaxy coevolution model should aim to quantitatively predict the differences in the host-galaxy properties for type~1 and type~2 AGNs, and these can help us to constrain the coevolution model more effectively.
}
\end{itemize}

\acknowledgments
Acknowledgments. We acknowledge the \texttt{CIGALE} team, from which we obtained many valuable suggestions on using \texttt{CIGALE} to conduct SED fitting. FZ and YX acknowledge support from the 973 Program (2015CB857004), NSFC-11890693, NSFC-11473026, NSFC-11421303, the CAS Frontier Science Key Research Program (QYZDJ-SSW-SLH006), and the K.C. Wong Education Foundation. GY and WNB acknowledge Chandra \mbox{X-ray} Center grant AR8-19011X and NASA ADP grant 80NSSC18K0878.

\bibliography{HostType}

\end{document}